%% file: ms.tex
\documentclass[12pt,preprint]{aastex}
\newcommand     \etl    {et al.}
\shortauthors{Bai \etl}
\shorttitle{Star Formation in MS 1054-03}

\begin{document}
\title{IR observations of MS 1054-03: Star Formation and its Evolution in Rich Galaxy Clusters}
\author{Lei\,Bai\altaffilmark{1},
Delphine\,Marcillac\altaffilmark{1},
George\,H.~Rieke\altaffilmark{1},
Marcia\,J.~Rieke\altaffilmark{1},
Kim-Vy\, H.~Tran\altaffilmark{2,3,4,5}
Joannah\,L.~Hinz\altaffilmark{1},
Gregory\, Rudnick\altaffilmark{6},
Douglas\,M.~Kelly\altaffilmark{1},
and Myra~Blaylock\altaffilmark{1}}
\email{bail@as.arizona.edu}
\altaffiltext{1}{Steward Observatory, University of Arizona, 933 N. Cherry Avenue, Tucson, AZ 85721}
\altaffiltext{2}{NSF Astronomy \& Astrophysics Fellow}
\altaffiltext{3}{NOVA Fellow}
\altaffiltext{4}{Harvard-Smithsonian Center for Astrophysics, 60
  Garden Street, Cambridge, MA 02138}
\altaffiltext{5}{Leiden Observatory, Leiden University, Niels Bohrweg
  2, 2333 CA Leiden, The Netherlands}
\altaffiltext{6}{National Optical Astronomy Observatory, 950 North Cherry Avenue, Tucson, AZ 85719; Leo Goldberg Fellow}
\begin{abstract}
We study the infrared (IR) properties of galaxies in the cluster MS 1054-03 at $z=0.83$ by combining MIPS 24 \micron\ data with spectra of more than 400 galaxies and a very deep K-band selected catalog. 
19 IR cluster members are selected spectroscopically, and an additional 15 are selected by their photometric redshifts.
We derive the IR luminosity function of the cluster and find strong evolution compared to the similar-mass Coma cluster.
The best fitting Schechter function gives $L^{*}_{IR}=11.49^{+0.30}_{-0.29}L_{\sun}$ with a fixed faint end slope, about one order of magnitude larger than that in Coma.
The rate of evolution of the IR luminosity from Coma to MS 1054-03 is consistent with that found in field galaxies, and it suggests that some internal mechanism, e.g., the consumption of the gas fuel, is responsible for the general decline of the cosmic star formation rate (SFR) in different environments.
The mass-normalized integrated SFR within 0.5$R_{200}$ in MS 1054-03 also shows evolution compared with other rich clusters at lower redshifts, but the trend is less conclusive if the mass selection effect is considered. 
A nonnegligible fraction ($13\pm3$\%) of cluster members, are forming stars actively and the overdensity of IR galaxies is about 20 compared to the field.
It is unlikely that clusters only passively accrete star forming galaxies from the surrounding fields and have their star formation quenched quickly afterward; instead, many cluster galaxies still have large amounts of gas, and their star formation may be enhanced by the interaction with the cluster.
\end{abstract}
\section{Introduction}
It has long been known that galaxies in rich clusters are quite different from those in the field.
A lower star formation rate (SFR) in cluster galaxies compared with the field at the same redshift has been found from local epochs up to $z\sim 1$ \citep{Kennicutt83,Balogh97,Hashimoto98,Poggianti99,Postman01,Gomez03}.
Several physical mechanisms are proposed to explain this suppression effect, e.g., galaxy-galaxy interaction, ram pressure stripping, and strangulation of the gas reservoir of galaxies \citep[see][and references therein]{Boselli06}.
Although all of those processes may play a role in suppressing the SFR in clusters, which one leads to the major effect is still controversial \citep{Balogh97,Hashimoto98,Poggianti99,Lewis02,Balogh04,Kauffmann04}.
More detailed study of the star forming activities in rich clusters is necessary to clarify the environmental effects on the SFR.

Aside from the general suppression compared with the field, the star formation activity of galaxies in clusters also evolves with epoch.  
This behavior was first discovered as the increasing number of blue galaxies in high $z$ clusters compared to local ones, the so-called Butcher-Oemler effect \citep{BO78, BO84}.
Spectroscopic studies \citep{Dressler82,Dressler83, Ellingson01, Tran05} reveal those blue galaxies as star-forming galaxies and indicate increased star forming activity in distant clusters compared to their relatively 'quiet' local kin.
In parallel to this trend in clusters, the average cosmic star formation rate has also experienced a rapid decline since $z\sim 1$ \citep{Lilly96,Madau98,Hopkins04,LeFloch05,Pablo05}.
This raises interesting questions of what drives the cosmic decline and how it relates to the decline of the SFR in clusters.
If clusters only passively accrete star forming galaxies from the field, then the increased number of blue galaxies in high $z$ clusters may merely result from the increased number of blue galaxies in the field and clusters would show a similar evolution in SFR as in the field. 
Also, if the cosmic decline of the SFR is mainly caused by an overall progressive gas consumption in galaxies, the same evolutionary trend in different environments will be a natural consequence.
On the other hand, if the SFR suppression is largely caused by an environment-dependent mechanism, the rate of decline of the cosmic SFR is expected to be correlated with the hierarchical growth of structure.

These questions may lack straight forward answers in the sense that all of the possible mechanisms may play a role in shaping the SFR of galaxies we observe, but with importance varying with epoch and environment.
However, an effort to measure the SFR in clusters at different redshifts \citep[e.g.,][]{Poggianti06} and a comparison of clusters with low density regions can provide us more insights into these issues.

Most of the previous studies of star forming activity in rich clusters are based on the galaxy colors, UV luminosity and emission-line measurement.
The colors of galaxies depend on the star formation history, as well as reddening, initial mass function (IMF), age, and metallicity.
They can only provide an approximate estimate of the star formation history of the galaxies.
The UV continuum directly traces the emission of the young stellar population ($<10^8$ yr), but it is heavily attenuated by dust and presents only the unobscured star formation, which is usually a small fraction of the total star formation in galaxies \citep[e.g.,][]{Buat05}. 
Several emission lines, e.g., $H{\alpha}$, [O II]$\lambda3727$, provide sensitive, instantaneous measurement of star formation (SF).  
The spectroscopic surveys of $H{\alpha}$ lines are mostly limited to galaxies with $z<0.4$, and it is hard to extend them to a large number of galaxies. 
The $H{\alpha}$ narrow band imaging technique \citep[e.g.,][]{Kennicutt83, Gavazzi98, Balogh00, Finn04} is more efficient but it requires specific filters for different redshifts and the continuum measurement could bring extra uncertainties.
Due to the [O II]/$H{\alpha}$ variations in galaxies and the higher extinction, the [O II]$\lambda3727$ lines are a less precise SFR tracer than $H{\alpha}$, but they can be applied to higher redshift ($z>0.4$) galaxies. 
Both $H{\alpha}$ and [O II]$\lambda3727$ emission line measurements are affected by dust extinction, which can be very high for strongly star-forming galaxies \citep[e.g.,][]{Choi06}.
As a result, robust measurements of the SFR unaffected by extinction in clusters are still lacking, especially at high redshift ($z>0.5$).

On the other hand, the infrared (IR) bolometric luminosity from the interstellar dust heated by the young stars in galaxies can provide a sensitive SFR indicator minimally affected by extinction \citep{Kennicutt98}.
Moreover, because mid-IR broadband emission shows a good correlation with the total IR luminosity \citep[e.g.,][]{Takeuchi05, Dale05}, it has become common practice to use single band mid-IR emission as a star formation indicator.
The mid-IR correlates well over a large luminosity range with the extinction corrected optical and near infrared SFR indicators \citep[e.g.,][]{Alonso06}.
Although there are many uncertainties involved in this method, e.g., the escape of the UV photons in optically thin regions, the heating of dust due to the ionizing photons from older stellar populations and the uncertainties in converting single band emission into the total IR luminosity \citep[e.g.,][]{Pablo06}, it is still a robust method to measure obscured SF in luminous galaxies. 

With data from the European Space Agency's Infrared Space Observatory (ISO), the IR-bright galaxies in quite a few clusters have been studied, with many important results \citep[see review in ][]{Metcalfe05}. 
In their study of A1689, \citet{Duc02} found at least 30\% of the 15 \micron\ sources show no evidence of current star formation in the optical spectra and that at least 90\% of the SF in A1689 is obscured by dust.
Their study shows the importance of using mid-IR emission to detect obscured SF in clusters. 
However, with only a small sample of low to mid-redshift clusters ($z<0.5$) studied by ISO, a complete understanding of the SF in clusters has not yet been achieved, especially for high redshift clusters.

The Multiband Imaging Photometer \citep[MIPS,][]{Rieke04} on the $Spitzer$ Space Telescope, with its high resolution and sensitivity, provides efficient measurements of IR emission for large samples of galaxies.
Using the 24 \micron\ observations by MIPS, we can study the SFR of clusters up to $z\sim1$.
These data provide a good chance to expand and explore the study of SF in clusters, leading to systematic and comprehensive understanding of galaxy evolution in them.
We have already studied the IR properties the Coma cluster \citep{Bai06} at $z=0.02$ and RXJ0152.7-1357 (RXJ0152 hereafter) \citep{Marcillac07} at $z=0.83$.
Here, we present an IR study of another well known cluster MS 1054-03 ($z=0.83$).
At this redshift, MIPS 24 \micron\ data trace the rest frame 13 \micron\ flux.
The 12-15 \micron\ emission has been demonstrated particularly by \citet{Roussel01,Takeuchi05,Alonso06} to be an accurate SF tracer. 
In \S 2, we describe the MIPS data for the cluster.
\S 3 discusses the analysis of the IR and optical data. 
We present the IR luminosity and additional IR properties of the cluster in \S 4.
In \S 5, we discuss the results and we summarize them in \S 6.
Throughout this paper, we assume a $\Lambda$CDM cosmology with parameter set $(h,\Omega_{0},\Lambda_{0}) = (0.7,0.3,0.7)$.

\section{DATA}
\subsection{Observations}
The MS 1054-03 field was observed at 24 \micron\ by MIPS on June 2005 in photometry mode.
The total MIPS field has a size of $5\arcmin \times 10\arcmin$.
The integration time is about 3600 second pixel$^{-1}$ in the central $5\arcmin \times 5\arcmin$ region, and is about 1200 second pixel$^{-1}$ in the rest of the field.
The data were processed with the MIPS instrument team Data Analysis Tool \citep{Gordon05,Engelbracht07} and scan-mirror-dependent flats were used.
The final mosaic has a pixel scale of $\sim 1.25\arcsec$ pixel$^{-1}$ and a point-spread function (PSF) with FWHM $\sim 6\arcsec$.

The spectroscopic data were obtained with the Low Resolution Imaging Spectrograph on the Keck Telescope based on the wide-field images taken by the $Hubble ~Space~ Telescope ~(HST)$ WFPC2 in F606W and F814W \citep{vanDokkum00,Tran07}.
They cover the $5\arcmin \times 5\arcmin$ central region of the cluster and yield more than 300 redshifts, adding to a total of more than 400 galaxies with previously known redshifts.
Among them, 144 sources with reliable redshifts are identified as cluster members.
The spectroscopic survey is about 50\% complete down to $I_{814}=22$ mag and it is mostly coincident with the central deep region of the 24 \micron\ observations.

Along with the HST photometric data, $UBV$ and near-IR $J_{s}HK_{s}$ data in a similar region of this field were obtained with FORS1 and ISAAC on the Very Large Telescope (VLT) as the part of the Faint InfraRed Extragalactic Survey (FIRES).
From these data a K-band selected catalog with 1859 sources was extracted.
The photometric catalog is presented in \citet{Forster06} and it is 90\% complete to $K_{s,AB} \approx 24.1$ mag.
Photometric redshifts ($z_{ph}$) were derived from this catalog using the method presented by \citet{Rudnick01,Rudnick03}.
The photometric redshifts are less accurate than the spectroscopic redshifts ($z_{sp}$), with $\delta z/(1+z_{sp}) = 0.074$ for $z < 1$, but the FIRES catalog is much deeper than the spectroscopic data.
Therefore, we use photometric redshifts to supplement the spectroscopic catalog.

\subsection{Source Detection and Completeness}
Since the cluster members are not resolved at 24 \micron, we used DAOPHOT II \citep{Stetson87}, a package for PSF fitting photometry, to detect sources and measure their fluxes. 
We follow the same strategy as described in \citet{Papovich04}.
Because of the significant zodiacal IR emission at low ecliptic latitudes, the 24 \micron\ background level in the MS 1054-03 field is fairly high, averaged at about 40 MJy sr$^{-1}$.
Even with more than 3000 seconds of integration time, the detection limit is not as deep as in some low background regions.
By adding artificial point sources into the image, we found a 80\% completeness limit at about 80 $\mu$Jy. 
In the left panel of Fig.~\ref{f_complete}, we plot the completeness of the 24 \micron\ detections $vs.$ the flux density.
The completeness drops from 80\% at about 80 $\mu$Jy to only 50\% at 50 $\mu$Jy.
We detected about 240 sources with $f_{24}>50~\mu$Jy in the central region that is covered both by IR and optical data, about 180 of them with $f_{24}>80~\mu$Jy.

\section{ANALYSIS}
\subsection{Spectroscopically Confirmed Cluster Members}
We select the galaxies with $0.81 < z_{sp} < 0.85$ as cluster members, which corresponds to a 3-$\sigma$ line-of-sight velocity dispersion of $1156\pm82$ km s$^{-1}$ \citep{Tran07}.
We select 144 cluster members out of around 400 galaxies with spectroscopic data.
Since the spectroscopic data are only 80\% complete down to $I_{814}=21$, we need to correct for the incompleteness to avoid bias.
In the right panel of Fig.~\ref{f_complete}, we plot the ratio of galaxies with successful spectroscopic redshift measurements among all galaxies in the imaging data as a function of $I_{814}$.
We use this curve to correct for the spectroscopic incompleteness when calculating the IR luminosity function (LF). 

\subsection{Photometrically Identified Candidate Cluster Members}
Even though we can roughly correct for the spectroscopic incompleteness of our IR sample using the completeness curve, about one third of the IR luminous members have a magnitude of $I_{814}>22$, where the incompleteness is larger than 50\% and a simple correction can be erroneous.
In addition, the $I_{814}$ band selection is biased toward the blue galaxies and may miss some dusty star forming galaxies with extreme red colors. 
This would make the incompleteness correction based on the $I_{814}$ magnitudes inadequate.
Therefore, we also use photometric redshifts to help select the cluster member sample.

Because most of the cluster members have $I_{814}-K_{s,AB} \approx 1.0$, the 90\% completeness limit ($K_{s,AB}\approx 24.1$) of the photometric survey corresponds to $I_{814} \approx 25.1$, indicated as a dashed vertical line in the right panel of Fig.~\ref{f_complete}.
This limit is about three magnitudes deeper than the spectroscopic survey.
However, the uncertainties in the photometric redshifts are large ($\delta z \sim 0.14$ at the cluster redshift $z_{cl}=0.83$) compared to those of the spectroscopic redshifts and their distribution is non-Gaussian.
Because of this, a simple cut in $z_{ph}$ is not effective to select cluster members and would cause large contamination. 
Therefore, we use the probability curve of the $z_{ph}$ deduced from Monte-Carlo simulations \citep{Rudnick03} to select the cluster members.
If the integrated probability of cluster membership for a galaxy with $K_{s,AB} < 24.1$ over the range of $z_{cl}-0.14 < z < z_{cl}+0.14$ is larger than 60\% (normalized by the total probability), we designate this galaxy as a cluster member.
The 60\% threshold is selected to best balance between maximizing correct selections and meanwhile minimizing incorrect selections when applied to galaxies with spectroscopic measurements.
When the threshold is set to 70\%, the incorrect selection drops from 17\% to 15\%, but the correct selection also drops from 83\% to 73\%. 
Altogether, we select 454 candidate cluster members from 1858 sources in this region, three times more than the members selected by spectroscopic redshifts alone.

The performance of the photo-z selection is expected to decrease for faint $K$ band sources. 
As the uncertainties in the fluxes increase, the internal uncertainties in the photo-z increase too. 
Therefore, the probability curve of $z_{ph}$ broadens and its integrated value in the same redshift range will be lower.
As a result, more faint galaxies will be rejected given the same threshold.

\subsection{Crossmatch between Optical/NIR Sources and 24 \micron\ Sources}
In a crowded field such as MS 1054-03, we need to be careful in crossmatching between the Optical/NIR sources and the 24 \micron\ sources.
The optical and NIR images have an absolute astrometric accuracy of $< 0\farcs5$ \citep{Forster06}.
The astrometry of the 24 \micron\ image is calibrated using the USNO-B1.0 catalog \citep{Monet03}, and has an accuracy of rms $<0\farcs6$.
However, due to the rather large FWHM ($\sim 6\arcsec$) of the 24 \micron\ image, this accuracy in position can only be achieved for bright IR point sources, namely with a 24 \micron\ flux $> 100~\mu$Jy.
For faint sources, as shown by our simulations, the average uncertainties of the positions are about $1\arcsec$. 
We use a radius of $2\arcsec$ ($\sim  15$ kpc at $z=0.83$) to correlate the optical/NIR cluster members with their IR counterparts.
This matching radius accounts for the possible displacement between the optical/NIR and 24 \micron\ brightness centroids, the astrometric uncertainties and local astrometric offsets.
If more than one counterpart is found in this radius, the nearest one is picked.
We estimate the chance of random matches by randomly re-distributing the IR sources and matching them with the same criteria.
We found only $4.8\pm0.9\%$ random matches.
Using this criterion, we obtain the preliminary matching lists for both spectroscopic and photometric samples.
We then carefully check each individual source by eye to exclude any apparent mis-identification, e.g., contamination from nearby bright IR sources.
Finally, we identify 19 sources selected by spectroscopic redshift and an additional 15 sources selected by photometric redshift with IR emission $\geq 50 ~\mu$Jy.
We refer to those 19 IR galaxies with spectroscopic redshifts as our spectroscopic sample, and  those 19 sources plus 15 sources selected by photometric redshifts as our combined sample. 
For the 15 photometrically selected galaxies, we will use the cluster redshift as their redshifts in the following study.
In addition, for most of these IR galaxies, we use the IRAC data kindly provided by the FIRES group to confirm the crossmatching.

\subsection{Incompleteness Correction}
\begin{figure}
\epsscale{1.0}
\plotone{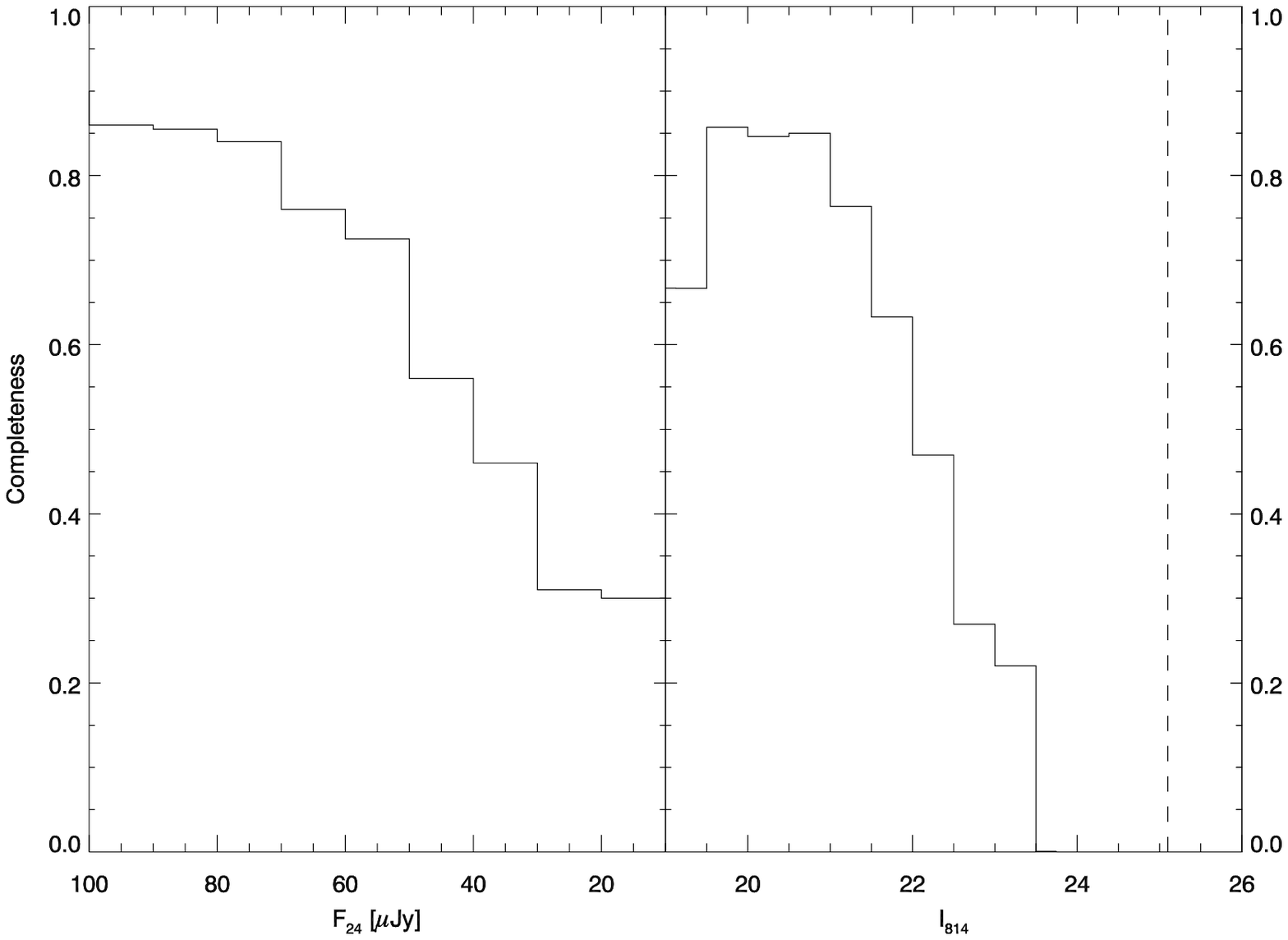}
\caption{The completeness of the IR and spectroscopic surveys. The left panel is the completeness at 24 \micron. The sample is about 80\% complete down to $f_{24}\approx80~\mu$Jy. The right panel is the completeness of the spectroscopic survey as a function of $I_{814}$ magnitude. The dashed vertical line is approximately the 90\% completeness limit of the photometric survey.}
\label{f_complete}
\end{figure}

Many of the sources in our final samples are fainter than the 80\% completeness limit of the spectroscopic and 24 \micron\ surveys.
It is therefore necessary to correct for the incompleteness of both surveys to have an unbiased number count.
To do this, we use the inverse of the completeness curves in the 24 \micron\ and $I_{814}$ bands given in Fig.~\ref{f_complete} as the weighting functions to calculate LFs.
For the 19 galaxies in the spectroscopic sample, we correct both for the spectroscopic incompleteness and 24 \micron\ incompleteness according to each galaxy's $I_{814}$ magnitude and 24 \micron\ flux density. 
All the galaxies in the combined sample are brighter than the 90\% completeness limit of the photometric survey, so we only correct for the incompleteness in the 24 \micron\ detections.

The incompleteness correction can be very large (see Fig.~\ref{f_complete}), especially for faint galaxies in the spectroscopic sample. 
It boosts the number density up to 3 times at the faint end of the luminosity function. 
We will discuss the effect of the incompleteness correction in \S4.1. 

\subsection{Deduction of the Total IR Luminosity}
\begin{figure}
\epsscale{1.0}
\plotone{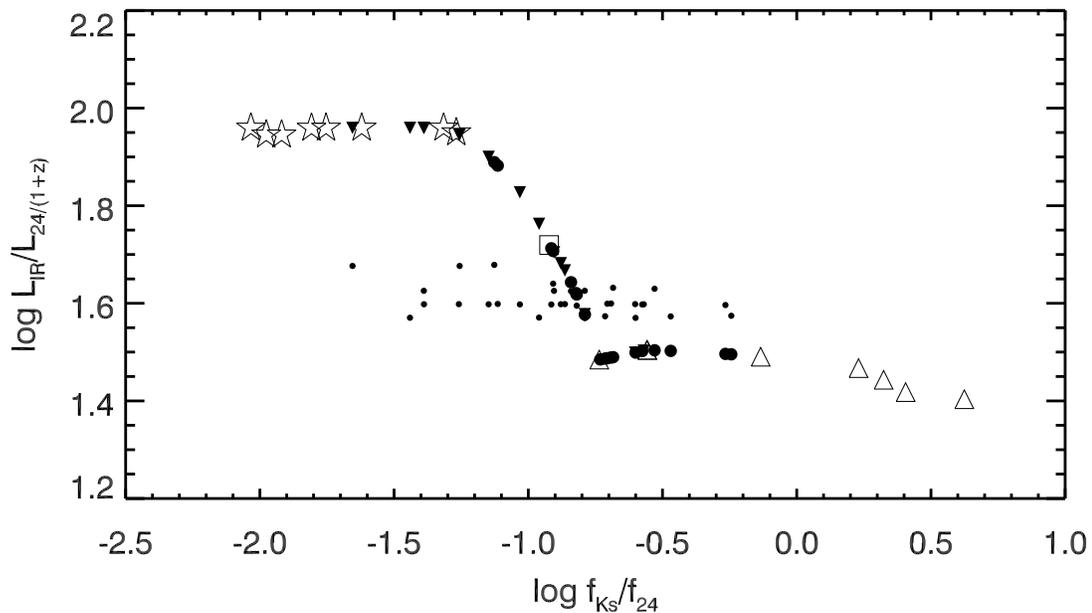}
\caption{The rest frame $L_{IR}/L_{24/(1+z)}$ ratio as a function of $K_{s}$ - 24 color. The open symbols are the data points deduced from the template SEDs given by \citet{Devriendt99}. Upward triangles, square, and stars denote the normal spirals, LIRG and ULIRGs. The filled circle and the downward triangle are the results of the interpolation of the galaxies of the spectroscopic and combined samples from their $K_{s}$ - 24 colors.
The small dots are the $L_{IR}/L_{24/(1+z)}$ ratio of galaxies deduced from the second method in \S 3.5.}
\label{f_ratio}
\end{figure}

To maintain continuity with \citet{Bai06}, we use their method to determine total IR luminosities.
We shift the SEDs given by \citet{Devriendt99} to the cluster redshift to deduce the rest-frame total IR luminosities ($L_{IR}, \lambda = 8 - 1000$ \micron) of the galaxies.
Those SEDs are based on a sample of nearby galaxies and include three types: normal spirals, luminous IR galaxies (LIRGs) and ultraluminous IR galaxies (ULIRGs).
The deduction of the total IR luminosity depends primarily on the ratio between the rest frame $L_{IR}$ and the 13 \micron\ luminosity ($L_{24/(1+z)}\sim L_{13}$) for each galaxy.
The template SEDs indicate that this ratio is almost constant within each type, but increases by three times from normal spirals to ULIRGs (see Fig.~\ref{f_ratio}).
Since the $K_{s}$ band (similar to rest frame $J$ band) flux and 24 \micron\ (similar to rest frame 13 \micron) flux are good indicators of the old and star-forming components of galaxies respectively, the color between these two bands ($f_{K_{s}}/f_{24}$) can be used to distinguish different types of galaxies.  
In Fig.~\ref{f_ratio}, we show the correlation between $L_{IR}/L_{24/(1+z)}$ and $f_{K_{s}}/f_{24}$ for each type of SED.
The open stars are ULIRGs, the square is a LIRG, and the open triangles are normal spirals.
We interpolate the $f_{K_{s}}/f_{24}$ colors of the cluster members linearly into the correlation given by template SEDs and get a $L_{IR}/L_{24/(1+z)}$ ratio for each galaxy.
The filled circles are the spectroscopic sample and the filled upside down triangles are the combined sample.
According to their $f_{K_{s}}/f_{24}$ colors, about one third of the galaxies in our spectroscopic sample and about half in the combined sample have ULIRG or LIRG SEDs.
Using the $L_{IR}/L_{24/(1+z)}$ ratio given by the interpolation, we deduce the total IR luminosity from the 24 \micron\ flux of each galaxy in our samples (see Table 1).

The method we used above basically assumes that there is no intrinsic variation in SEDs for galaxies with same $f_{K_{s}}/f_{24}$ colors and that the templates represent a complete sample of IR galaxies up to $z\sim 0.8$. 
However, both of these assumptions are questionable, given the large variation of IR SEDs among star forming galaxies and the possible evolution of galaxy properties from $z=0.8$ to $z=0$. 
To estimate the uncertainties of $L_{IR}$ caused by the limitations of the SED templates, we used a different set of SEDs from \citet{Dale02} and a strategy described in \citet{Marcillac06b} to deduce the total IR luminosities. 
\citet{Marcillac07} use this method to deduce the total IR luminosities in another cluster, RXJ0152, at a similar redshift.
We plot the deduced $L_{IR}/L_{24/(1+z)}$ ratio of galaxies vs. $f_{K_{s}}/f_{24}$ colors as small dots in Fig.~\ref{f_ratio} for comparison, though the analysis itself does not depend on $f_{K_{s}}/f_{24}$ color.
This method gives a slightly smaller typical $L_{IR}$ compared to the first method, by a factor of $0.9\pm0.3$ on average.
The difference is more pronounced for those galaxies with a smaller $f_{K_{s}}/f_{24}$ color ($<-1$), where the difference is up to a factor of 2, and may be caused by the mis-classification of SED types with only one color.
To exclude this possibility, we compared the multi-wavelength photometry of sample galaxies (optical + NIR + IRAC + MIPS 24 \micron\ ) with the model SEDs from \citet{Devriendt99} and confirmed they do have LIRG/ULIRG type SEDs.
The difference, caused by the wide SED variations from galaxy to galaxy, is typical of methods to estimate total IR luminosities from 24 \micron\ measurements \citep[e.g.,][]{Papovich02,Dale05}. 
It does not affect the results of this paper significantly.
If we do not consider the uncertainties caused by the SED fitting, the error of $L_{IR}$ is dominated by the flux uncertainties at 24 \micron, which are typically 50\% for the galaxies studied in this paper. 

\subsection{Contamination from AGNs}
When we deduce total IR luminosities for the galaxies, we assume their IR emission is entirely from emission by dust heated by star forming activity, neglecting the possible contribution from active galactic nuclei (AGNs).
Although optical studies suggest that AGNs reside in only about one percent of galaxies in clusters up to $z\sim0.5$ \citep{Dressler99}, recent X-ray surveys have found an excess of point sources in cluster fields, many of which are confirmed as cluster AGNs \citep[e.g.,][]{Martini02}.
These discoveries suggest that AGN contamination may be an issue, especially for our small number samples.

In the MS 1054-03 field, surveys in the radio and the X-ray bands have been analyzed to identify the possible AGNs. 
\citet{Best02} conducted an extremely deep 5-GHz radio observation and found 34 radio sources, 8 of which are confirmed as cluster members by their spectroscopic data.
On the basis of the [\ion{O}{2}] emission line flux and radio flux density ratio, they further conclude that 6 of these 8 radio sources are AGNs, one source (No 5) is a star-forming galaxy and one (No 14) is ambiguous.
\citet{Johnson03} analyzed the 91 ks {\it Chandra} observations of the cluster and detected 47 X-ray sources.
Among them, two sources are confirmed as AGNs (source 7 and source 19); source 19 is also detected in the radio.
Altogether, there are 8 possible AGN members in the cluster.
To avoid losing possible AGN candidates in our IR galaxy sample due to the incompleteness of the spectroscopic survey, we crossmatched the IR galaxies with all the 34 radio sources and the 47 X-ray sources.
The two confirmed AGNs (X-ray source 7 and source 19) and the No 5 radio source are detected in the IR.
We exclude the two AGNs from both of our samples.
For the No 5 radio source, we use a radio spectral index of -0.8 and the formula given by \citet{Hopkins03} to convert the radio flux to the SFR. 
We deduce a SFR of about 88 $M_{\sun}~{\rm yr^{-1}}$ from its radio flux, which is consistent with the SFR estimated from the total IR luminosity, $\sim 61~ M_{\sun} {\rm yr^{-1}}$, using the conversion formula given by \citet{Kennicutt98}.
This agreement further confirms radio source No 5 as a star-forming galaxy.
Due to the limitations of the spectroscopic data and the sensitivity of the X-ray survey, we can not totally exclude all AGN contamination from our IR galaxy samples, but the fact that only one star forming IR galaxy in our sample is detected either in the radio or in the X-ray band indicates the contamination is negligible.

\subsection{Comparison of the IR and [\ion{O}{2}] emission line deduced SFRs}
\begin{figure}
\epsscale{1.0}
\plotone{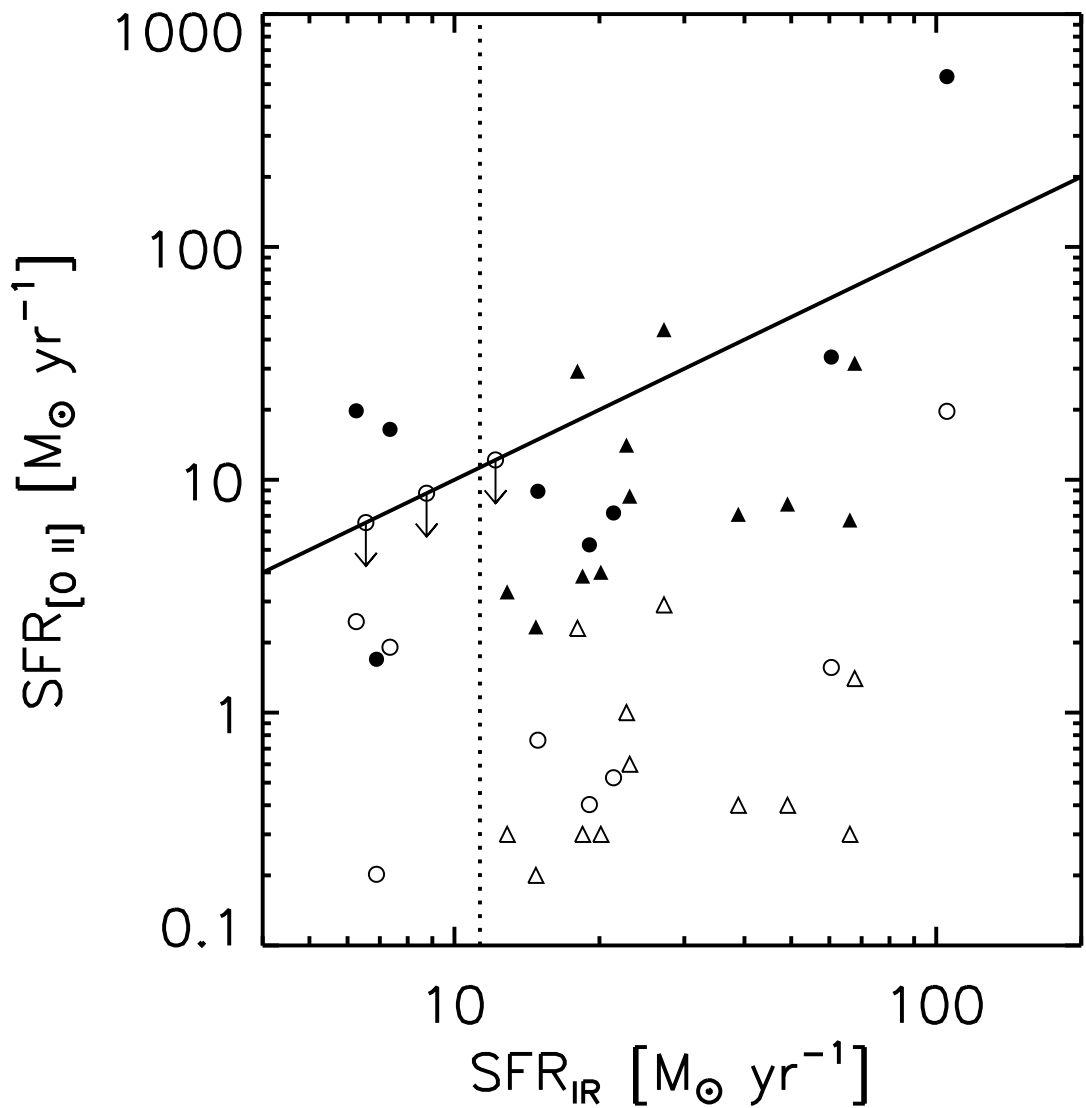}
\caption{The comparison of SFR$_{IR}$ and SFR$_{[OII]}$.
The filled and open circles are the SFRs of the IR galaxies in MS 1054-03, with and without extinction correction for SFR$_{[OII]}$.
For the three IR galaxies without [\ion{O}{2}] emission lines, SFR$_{IR}$ are used as the upper limit of SFR$_{[OII]}$ and they are plotted as open circles with downward arrows.
The filled and open triangles are the SFRs of the IR galaxies in RXJ0152, with and without extinction correction for SFR$_{[OII]}$.
The dotted vertical line is the SFR approximately corresponding to the 80\% completeness limit of the 24 \micron\ observation. 
The solid line indicates the one-to-one correlation.}
\label{f_sfr_OII}
\end{figure}

Many previous studies of the SFR in clusters at $z>0.4$ rely on the [\ion{O}{2}] emission line as an indicator, which is very sensitive to extinction and metallicity.
We compare the SFR deduced from the [\ion{O}{2}] emission line luminosity and that from the total IR luminosity in Fig.~\ref{f_sfr_OII} for the IR galaxies in the spectroscopic sample.
Among 15 IR galaxies with [\ion{O}{2}] data, 12 galaxies have emission lines.
The [\ion{O}{2}] emission line luminosity is estimated by multiplying the equivalent width of the emission line by the continuum flux.
The continuum flux at the rest frame of the [\ion{O}{2}] line is approximated by the continuum flux in $V_{606}$.
The [\ion{O}{2}] emission line luminosity is converted to a SFR using the formula SFR$_{[\rm OII]}=(6.58\pm1.65)\times 10^{-42}{\rm L}_{[\rm OII]} ({\rm ergs~s}^{-1})$ \citep{Kewley04}, where L$_{[\rm OII]}$ is the luminosity corrected for extinction.
Without any extinction correction, the SFR$_{[\rm OII]}$ is smaller than the SFR$_{\rm IR}$ by more than one dex on average (the open circles), but with large scatter.
Since we do not have enough optical data to deduce extinction, we used the IR-luminosity-dependent extinction $A^{\rm IR}_{V}=0.75 {\rm log}(L_{\rm IR}/L_{\sun})-6.35$ mag given by \citet{Choi06} to correct for the dust attenuation.
This extinction formula is deduced from the ratio of the $SFR_{IR}$ and the SFR measured from emission lines, assuming $SFR_{IR}$ approximate the true SFR.
The galaxies in our sample all have a $A^{\rm IR}_{V}$ greater than 1.5.
We convert $A^{\rm IR}_{V}$ to the extinction of [\ion{O}{2}] line using the reddening curve of \citet{Calzetti00}, the same one used by \citet{Choi06}.
The extinction-corrected SFR$_{[\rm OII]}$ agrees with the SFR$_{\rm IR}$ reasonably well (the filled circles), with a scatter of about 0.5 dex.

For the three IR galaxies without emission lines, we plot the SFR$_{\rm IR}$ as the upper limits of their SFR$_{[\rm OII]}$ (the open circles with downward arrows).
There are also about a dozen [\ion{O}{2}] emission line galaxies (EW$_{[\rm OII]} > 5 \rm {\AA}$) not detected at 24 \micron.
For these galaxies, the lack of IR emission probably suggests relatively less dust and smaller extinction, so we used a fixed $A_{V}=1.0$ to deduce their SFR$_{[\rm OII]}$.
Their SFR$_{[\rm OII]}$ are all quite small, with a maximum value of 7 $M_{\sun}$ yr$^{-1}$, well below the 80\% detection completeness limit.

We also compared the SFR$_{[\rm IR]}$ of galaxies in RXJ0152 with the SFR$_{[\rm OII]}$ given by \citet{Homeier05}, correcting for extinction with $A^{\rm IR}_{V}$.
The open and filled triangles in Fig.~\ref{f_sfr_OII} are the data without and with extinction correction.
Again, the extinction corrected SFR$_{[\rm OII]}$ shows a better consistency with the SFR$_{\rm IR}$.
This agrees with the results of \citet{Marcillac07}, who also found a large amount of dust-embedded SF in RXJ0152.

The median values of $A^{\rm IR}_{V}$ for the IR galaxies with [\ion{O}{2}] emission lines in MS 1054-03 and in RXJ0152 are both about 2, corresponding to a correction factor of $\sim 14$ for the SFR$_{[\rm OII]}$.
This result implies the star-forming galaxies in these clusters are enshrouded heavily by dust and the SFR$_{[\rm OII]}$ without extinction correction only measures a small portion of the total SFR.
Even with the widely adopted extinction of 1 mag at H${\alpha}$ ($A_{V}\sim 1.2$), the emission line SFRs still underestimate the SFR by a factor of 4 for these IR bright galaxies.

\subsection{Comparison of the IR and Ultraviolet continuum deduced SFRs}
UV luminosity is also widely used to estimate the SFR of galaxies.
Although it is very sensitive to the dust extinction, it gives us access to the "non dusty" star formation and is therefore complementary to the IR-deduced SFR.
For the cluster members, we have derived the rest-frame $2200 \rm{\AA}$ luminosity ($L_{\nu,2200\AA}$) from the galactic extinction corrected $U-K$ photometry from \citet{Forster06}, using the methodology presented in \citet{Rudnick03}.
We estimate the SFR from $L_{\nu,2200\AA}$ using the formula given by \citet{Kennicutt98}.
The conversion assumes a Salpeter IMF and a constant SFR, with UV emission dominated by a stellar population younger than 100 Myr.
These assumptions are consistent with those used to deduce the IR SFR conversion formula we adopt in this paper \citep{Kennicutt98}. 

\begin{figure}
\epsscale{1.0}
\plotone{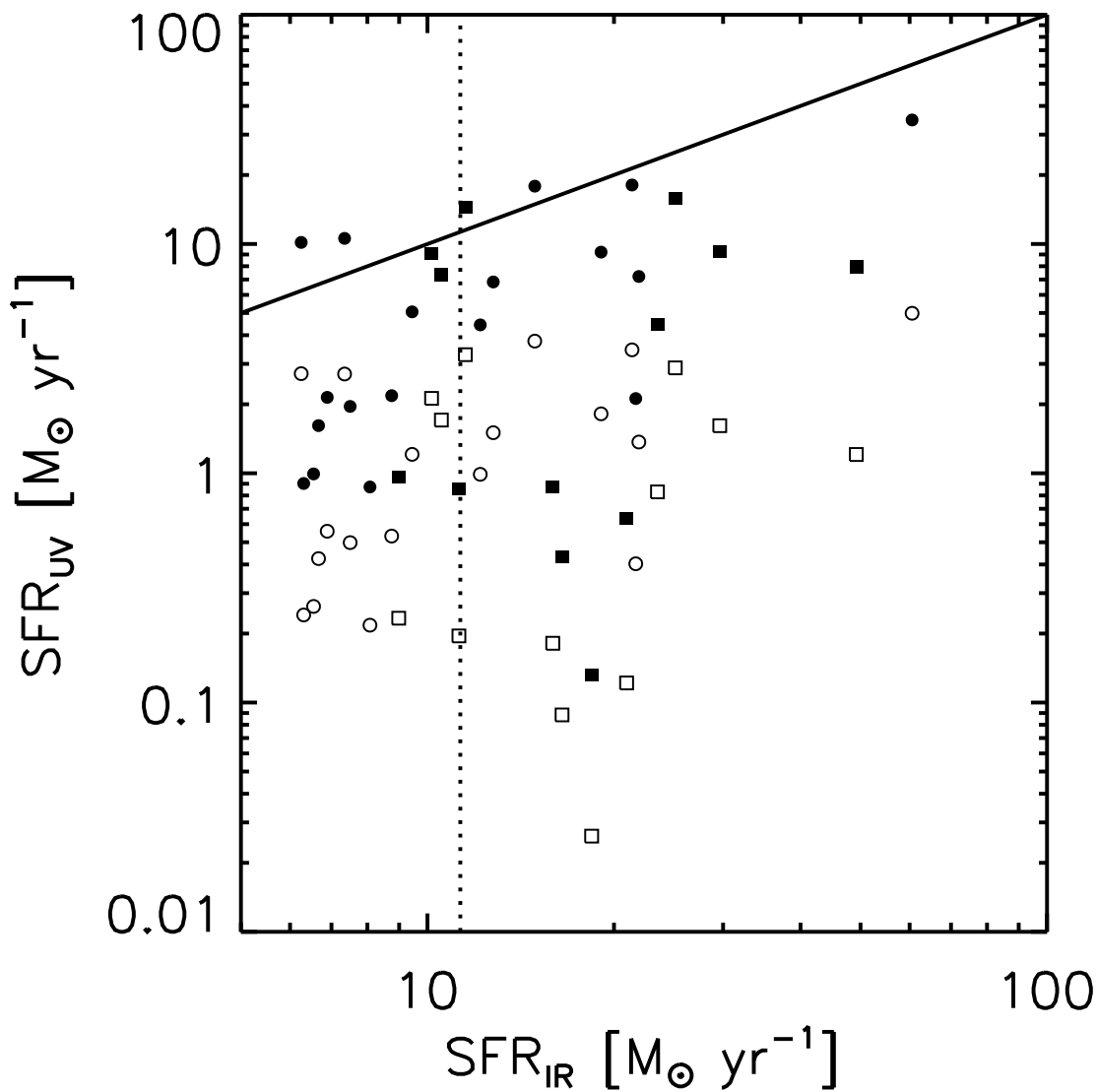}
\caption{The comparison of SFR$_{IR}$ and SFR$_{UV}$.
The open circles are the spectroscopically confirmed IR cluster members, and the open squares are the ones selected by their photometric redshifts.
The filled circles and squares are the results after applying extinction correction.
The dotted vertical line is the SFR approximately corresponding to the 80\% completeness limit of the 24 \micron\ observation.
The solid line indicates the one-to-one correlation.
}
\label{f_sfr_UV}
\end{figure}

In Fig.~\ref{f_sfr_UV},  we show the comparison between the UV continuum-deduced SFRs and the IR-deduced SFRs.
For the IR bright cluster members, the unobscured star formation are only a small fraction of the total star formation.
The median value of SFR$_{\rm IR}$/SFR$_{\rm UV}$ is $\sim$ 12 for the spectroscopic sample and $\sim$ 16 for the combined sample.
Such a large SFR$_{\rm IR}$/SFR$_{\rm UV}$ ratio is mainly due to the fact that our IR data is only sensitive to galaxies with $SFR \ga 10~ M_{\sun}$ yr$^{-1}$, where extinction is known to be large.
The combined sample has a larger SFR$_{\rm IR}$/SFR$_{\rm UV}$ ratio on average because the spectroscopic survey is $I$-band magnitude limited and is biased against the most dusty star forming galaxies.
If we calculate the visual gas medium extinction from $L_{IR}$ using the formula given by \citet{Choi06} (all the extinctions we mentioned in \S 3.7 are for the gas medium) and apply the corresponding UV stellar continuum extinction to SFR$_{\rm UV}$, we will have a better agreement between SFR$_{\rm IR}$ and SFR$_{\rm UV}$, as shown in Fig.~\ref{f_sfr_UV}.
However, even after this extinction correction, there are still many galaxies, especially the ones selected by photometric redshifts, showing a much smaller SFR$_{\rm UV}$ compared to SFR$_{\rm IR}$.

If we assume SFR$_{\rm IR}$ is the total SFR and directly estimate the NUV extinction by 2.5log(SFR$_{\rm IR}$/SFR$_{\rm UV}$), we will have a median $A_{NUV}$ of $2.7\pm0.5$ mag and $3.0\pm1.0$ mag for the spectroscopic sample and the combined sample respectively.
The high extinction we found here supports our assumption that SFR$_{\rm IR}$ provides a reasonable estimate of the total SFR.
The NUV extinction of the stellar continuum can be translated into the visual extinction of the gas medium using the reddening curve of \citet{Calzetti00}, $A_{V}=2.9\pm0.5$ mag and $A_{V}=3.3\pm1.1$ mag respectively for our two samples. 
\citet{Buat07} studied the extinction of a sample of LIRGs detected in the Chandara Deep Field South at $z=0.7$ using the ratio of the total IR and FUV luminosity.
They found an average FUV extinction of $3.33\pm0.08$ mag for their sample, corresponding to a gas medium visual extinction of $A_{V}=2.97\pm0.07$ mag 
\footnote{To be consistent with \citet{Choi06}, we use the reddening curve of \citet{Calzetti00} for all the extinction conversion in this paper, which indicates $A_{V}=4.05E(B-V)_{g}$.  
It is different from $A_{V}=3.1E(B-V)_{g}$ used by \citet{Buat07}.  
If we adopt their conversion method, the absolute value of the visual extinctions corresponding to the NUV and FUV extinctions will change, but the results of the comparison will remain the same.}
, which is in very good agreement with our results.

In addition to the IR cluster members, we also calculate the SFR$_{\rm UV}$ for all the cluster members (spectroscopic + photometric) without detectable IR emission.
Their SFR$_{\rm UV}$ are all at least two times smaller the detection limit of the SFR$_{\rm IR}$, which confirms there is no galaxy with a high level of star formation that is missed by IR selection due to the lack of dust.

\section{RESULTS}
\subsection{IR Luminosity Function}
After obtaining the total IR luminosity of each galaxy, we calculate the LF for each sample.
For the spectroscopic sample, we correct the number counts for the incompleteness in both the $I_{814}$ and 24 \micron\ bands.
We only correct for incompleteness in the 24 \micron\ detections for the combined sample.
The overlapping area between the spectroscopic survey and the 24 \micron\ observations is about 4.8 Mpc$^{2}$, and the overlapping area between the photometric survey and the 24 \micron\ observations is 5.5 Mpc$^{2}$. 

\begin{figure}
\epsscale{0.9}
\plotone{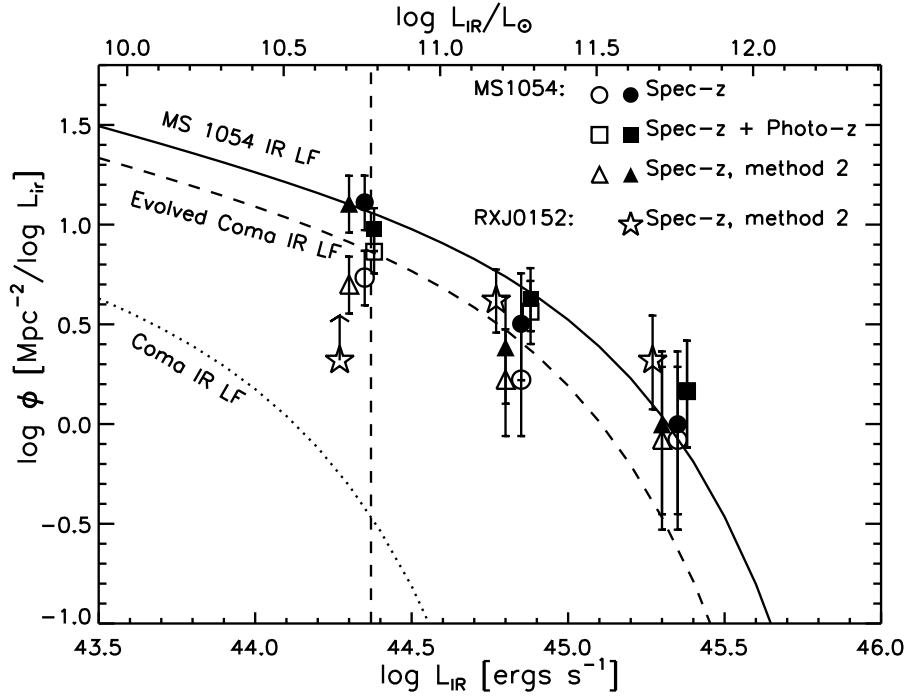}
\caption{The IR luminosity function of MS 1054-03. 
The open and filled circles are the result of the spectroscopic sample without and with spectroscopic and IR incompleteness correction.
The open and filled triangles are the result of the spectroscopic sample deduced from the second method in \S 3.5, without and with incompleteness correction. 
The open and filled squares (shifted to the bright end by 0.04 for clarity) are the result of the combined sample without and with IR incompleteness correction.
The open stars (shifted to the faint end by 0.04 for clarity) are the IR LF of RXJ0152 from \citet{Marcillac07}.
Since they only include galaxies with $f_{24} > 80 ~\mu$Jy and the data are very incomplete at log$L_{IR}< 44$, we draw the faintest point as a lower limit. 
The solid curve is the best fitting Schechter function to the corrected spectroscopic IR LF.
The dotted curve is the best fitting Schechter function to the IR LF of the Coma cluster.
The dashed curve is the Coma LF evolved to $z=0.83$ with the same evolution trend as the field IR LF. 
The vertical dashed line is the IR luminosity corresponding to the 80\% detection limit at 24 \micron.}
\label{f_LF}
\end{figure}

The IR LFs are shown in Fig.~\ref{f_LF}.
The open circles are the LF deduced from the spectroscopic sample without any incompleteness correction and the error bars are estimated by Poisson statistics \citep{Gehrels86}; the filled circles are the results corrected for the incompletenesses in both the spectroscopic and IR surveys.
The correction is quite significant except for the brightest data point.
The error for the incompleteness corrected LF is obtained by multiplying the original error by the incompleteness correction made at each data point.
Since we do not consider the error caused by the incompleteness estimate itself, the error bars should be considered to be lower limits to the actual errors.
Similarly, the incompleteness uncorrected and corrected data points of the combined sample are shown as open and filled squares respectively.
Even though the uncorrected LFs of the spectroscopic and combined samples exhibit a large difference, their incompleteness corrected ones agree with each other quite well.
This good agreement demonstrates that neither the simplified spectroscopic incompleteness correction nor the uncertainty of the photometric redshifts affects our resulting IR LFs significantly.
It also shows that there are few galaxies with extremely red optical-IR colors missed by the selection limit in the $I_{814}$ band. 

Despite this general agreement, the difference in the brightest data point may cause quite a large discrepancy when we try to fit the LF.
It also raises questions about the incompleteness correction because we expect it to be least significant for the brightest galaxies.
Two spectroscopically and two photometrically selected IR galaxies contribute to this data point.
The two galaxies selected by spectroscopic redshifts are both very bright ($I_{814}\approx20$) late type galaxies, and the two selected by photometric redshifts are both about two magnitudes fainter in the $I_{814}$ band and slightly brighter in the IR.
Even though the probability of the photometric redshifts of those two galaxies falling into the one sigma error range of the cluster redshift is more than 70\%, their best fitting photometric redshifts are both about 0.95.
As an independent check, P. G. P\'{e}rez-Gonz\'{a}lez helped us get another set of photometric redshifts for these two IR galaxies using a different fitting strategy \citep{Pablo05} and with IRAC photometric data as an addition.
These photometric redshifts have an average accuracy of $\Delta z = 0.08$.
The best fitting redshifts of those two galaxies are $0.97\pm{0.09}$ and $1.00\pm{0.11}$, both more than 1 $\sigma$ above the cluster redshift.
Their extremely large SFR$_{\rm IR}$/SFR$_{\rm UV}$ ratio, as shown in Fig.~\ref{f_sfr_UV}, also suggest them as background sources.
In addition, among the 20\% of photometric sources with spectroscopic data, one galaxy as bright as those two at 24 \micron\ is selected as a cluster member by its photometric redshift but shown to be a non-member by its spectroscopic redshift.
Statistically, it is possible that four more foreground or background contaminations may occur in the whole sample. 
However, we still can not rule out the possibility of those two sources as real cluster members given the uncertainties of the photometric redshifts.     
Spectroscopic data are needed to clarify the ambiguity. 

Because the incompleteness-corrected IR LF of the spectroscopic and combined sample are generally consistent, while the combined sample has larger uncertainties at the brightest end (the spectroscopic sample, on the other hand, should have the smallest uncertainties due to the incompleteness correction at this point), we select the incompleteness-corrected IR LF of the spectroscopic sample as the IR LF of the cluster.
We fit this LF with a Schechter function \citep{Schechter76}.
Since we only have three data points, we fix the faint end slope to the same value as the IR LF of the Coma cluster \citep{Bai06}.
We adopted a chi-square minimization method for the fitting. 
We also use the non-detection of the brighter galaxies beyond the brightest bin as a constraint during the fitting \citep{Bai06}.
The best fitted parameters are:
\begin{equation}
\alpha=1.41~ (fixed);~{\rm log}(L^{*}_{IR}/L_{\sun})=11.49^{+0.30}_{-0.29}.\
\end{equation}
The resulting fit is shown as the solid curve in Fig.~\ref{f_LF}.
Fitting the incompleteness-corrected IR LF of the combined sample gives an even larger $L_{IR}^{*}$, with ${\rm log}(L^{*}_{IR}/L_{\sun})=11.73^{+0.34}_{-0.23}$.
We only use the Poisson statistical errors for the fitting and do not consider the uncertainties caused by the errors in $L_{IR}$ estimation.  
The best-fitting parameters have large uncertainties, because the Schechter function fitting depends strongly on the brightest bin, which only includes two galaxies to constrain $L^{*}_{IR}$, and even small changes of the $L_{IR}$ of those galaxies can cause large changes in the best fitting parameters.
We also note that due to the degeneracy between the faint end slope and the characteristic IR luminosity ($L_{IR}^{*}$), the best-fitting $L_{IR}^{*}$ value we obtained here depends on the assumed faint end slope.
If we vary the faint end slope from its current value by $\pm0.2$, the best-fitting $L_{IR}^{*}$ would vary by $\pm0.13$.
However, by fixing the faint end slope and fitting to a Schechter function, we can quantify the difference between LFs.
Because no available infrared data in this redshift range penetrate significantly below the LIRG range, virtually all studies use a fixed low luminosity slope, including the field IR LF we compare with in this paper \citep{LeFloch05}.

We have tested the dependence of these fits on the uncertainties in the deduction of $L_{IR}$ with different methods.
If we use the $L_{IR}$ of the spectroscopic sample deduced from the second method listed in Sect. 3.5, the IR LF does not change significantly, as shown by open and filled triangles in Fig.~\ref{f_LF}. 
The best fitting function has a smaller ${\rm log}(L^{*}_{IR}/L_{\sun})=11.41^{+0.33}_{-0.52}$.
The difference is still within the one sigma Poisson error, suggesting that small number statistics dominate the uncertainty to define a best fitting LF, and the systematic error caused by different $L_{IR}$ deduction methods is negligible.
Therefore, in the rest of the paper, we only use the results from the first method.

\subsection{Comparison with Coma IR LF}
We compare the IR LF of MS 1054-03 to that of the Coma cluster, which has similar mass as MS 1054-03 \citep{Lokas03,Jee05a} and whose galaxy infrared luminosities are deduced using the same set of SEDs as the first method in this paper.
We plot the best fitted Schechter function of the Coma cluster IR LF as the dotted curve in Fig.~\ref{f_LF}.
The characteristic IR luminosity in MS 1054-03 is ten times larger than that of the Coma cluster (${\rm log}(L^{*}_{IR}/L_{\sun})=10.49^{+0.27}_{-0.24}$).
The surface density of the IR galaxies with ${\rm log}L_{IR} \geq 43.5$ expected from the MS 1054-03 LF is about 5 times larger than that in Coma.
We integrate the best fitted Schechter function of the MS 1054-03 IR LF from log$L_{IR}=44$ to log$L_{IR}=46$ and get a SFR density of 190 $M_{\sun}$ yr$^{-1}$ Mpc$^{-2}$, about 16 times larger than the SFR density of the Coma cluster ($\sim 11.4$ $M_{\sun}$ yr$^{-1}$ Mpc$^{-2}$). 

The significant difference between the IR LFs of MS 1054-03 and of Coma agrees with the general evolution trend found in the field IR LF \citep{LeFloch05,Pablo05}. 
\citet{LeFloch05} quantified the evolution of the IR LF in the CDF-S field in both density and luminosity as [$L^{*}_{IR} \propto (1+z)^{\alpha_{L}}, \phi^{*}_{IR}  \propto (1+z)^{\alpha_{D}}$], with the best fitting parameters $\alpha_{L}=3.15\pm1.6,~ \alpha_{D}=1.02\pm1.6$.
Corresponding to this, we estimate the difference of the two cluster IR LFs using the same parameters and get $\alpha_{L}=4.0^{+2.1}_{-2.2}, \alpha_{D}=1.4$.
For $\alpha_{D}$, we do not give an error estimate due to the large uncertainties of the best fitting $\phi^{*}_{IR}$ value.
The results agree within the errors and even suggest a slightly stronger evolution of these two cluster IR LFs compared to the field IR LF.
This is also demonstrated in Fig.~\ref{f_LF}, where the dashed curve corresponds to the LF of the Coma cluster evolved to $z=0.83$ using the field IR LF evolution law.
Both the incompleteness-corrected IR LFs of the spectroscopic and combined sample of MS 1054-03 fall above this curve.

The agreement between the evolution of the IR LF in these two clusters and the evolution in the field might suggest that the population of star forming galaxies in clusters is dominated by recently accreted field galaxies. 
Therefore, their IR LFs would not be very different from that of field galaxies and they would show similar evolution. 
However, this explanation is not favored by our following analysis (see \S 4.3).
More likely, the similarity in the evolution trend suggests that the cosmic SFR decline is caused by some general mechanism existing both in cluster and field environments, probably the consumption of the gas fuel for SF. 
Nevertheless, due to the intrinsic variation of cluster properties and with only two clusters in this comparison, these results are far from conclusive. 
More clusters need to be studied to confirm the evolution trend further. 

\subsection{Comparison with Field IR LF}
\begin{figure}
\epsscale{1.0}
\plotone{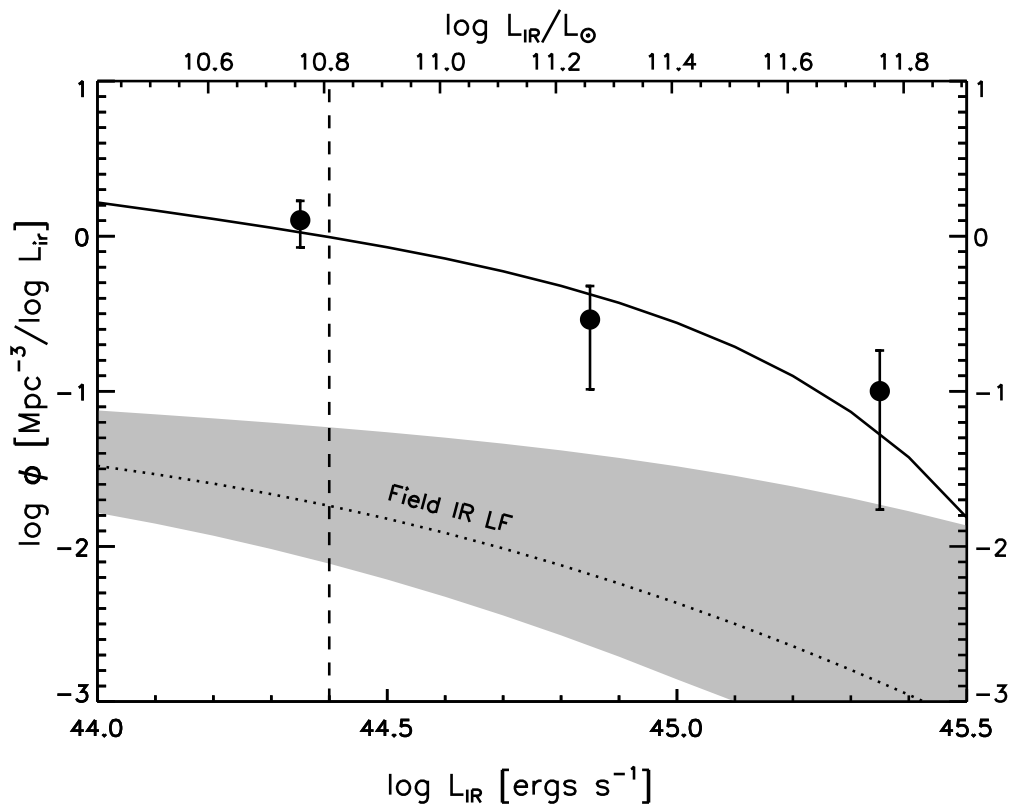}
\caption{The IR luminosity function of MS 1054-03 compared with the field IR luminosity function at $z\sim0.8$. The filled symbols are from the spectroscopic data after incompleteness correction. The solid curve is the best fitting Schechter function. The dotted curve is the field IR LF at the cluster redshift. The shaded area presents its one sigma uncertainties. The vertical dashed line is the IR luminosity corresponding to the 80\% detection limit at 24 \micron.}
\label{f_LF_comp}
\end{figure}

The IR LFs we deduced for the cluster are all projected LFs.
If we assume the cluster has a radius of about 5$R_{200}$ ($R_{200}$ is the radius within which the mean cluster density is 200 times the critical density of the universe at that redshift 
\footnote{$R_{200}$ is widely used as an approximation to $R_{virial}$. 
Strictly speaking, $R_{virial}$ is closer to $R_{100}$. 
The ratio of $R_{200}/R_{virial}$ depends on the mass distribution of a cluster.
For a NFW profile \citep{Navarro97} with $R_{200}/r_{s}=6$, R$_{200}$/R$_{virial}\sim0.75$.}
), we can calculate the IR LF per volume and compare it with the field IR LF from \citet{LeFloch05} at similar redshift.
5$R_{200}$ is near the turnaround radius of the cluster, and there will be few infalling galaxies beyond it.
Although the spectroscopic sample is almost free of contamination from field galaxies, the redshift selection ($0.81 < z <0.85$) still could include a few foreground and background field galaxies whose redshifts fall within the cluster velocity dispersion. 
To exclude this field contamination, before we convert the projected cluster IR LF to LF per volume, we calculate the projected field IR LF in a cylinder with a length corresponding to $z=0.81$ to $z=0.85$ and subtract it from the projected IR LF of the cluster.
In Fig.~\ref{f_LF_comp}, the filled circles are the incompleteness-corrected IR LF per volume of the spectroscopic sample after the field subtraction.
If we integrate the cluster LF in the range of $10^{10.8}<L_{IR}<10^{12}L_{\sun}$, it shows an overdensity of about 21 compared with the IR LF from the CDF-S field from \citet{LeFloch05}.
The overdensity calculated here can be affected by the cosmic variance from field to field, especially the CDF-S field, in which a lower galaxy density up to a factor of two is found compared with other fields \citep{Wolf03}.
However, as a first-order correction, \citet{LeFloch05} already normalized their IR LFs by the ratio between the $B$-band luminosity densities in the CDF-S and over the three fields of COMBO-17 \citep[Classifying Objects by Medium-Band Observations in 17 filters;][]{Wolf03,Wolf04,Bell04}.

Although the estimate of the actual value of the overdensity has some uncertainties, it is clear that there is an excess of IR galaxies in the cluster compared with the field. 
Such an excess of MIR sources is also found in RXJ0152 \citep{Marcillac07}, as well as in two intermediate redshift clusters Cl 0024+16 and MS 0451-03 (with a smaller significance) \citep{Geach06}.
Although the cluster shows a clear overdensity of the IR galaxies compared with the field, it is still smaller than the overdensity of the cluster in the optical bands. 
A fairer test to examine the star formation level in different environments is to compare the fraction of IR galaxies in the cluster and in the field.
Among the 144 spectroscopically confirmed cluster members, 19 have 24 \micron\ emission brighter than $50~ \mu$Jy and 6 of them have $L_{IR} > 10^{11} L_{\sun}$.
There are two AGNs also with 24 \micron\ emission, but we already excluded them from the sample.
Therefore, the fraction of star-forming galaxies with $f_{24} > 50 ~\mu$Jy in the cluster is about $13\pm3$\% and the fraction for LIRGs is $4\pm2$\%.
These fractions are barely affected by the incompleteness of the spectroscopic survey.
There are 211 cluster member candidates selected by the photometric redshifts with $K_{s,AB}<22$ (approximately the detection limit of the spectroscopic survey of $I_{814}=23$) and $15\pm3$\% of them are IR bright and $5\pm2$\% are LIRGs.
The results are therefore consistent with the fractions based only on the spectroscopically selected cluster members.
For the CDF-S field, we select the galaxies with $0.81<z_{ph}<0.85$ and $R<22.6$ mag using the photometric redshifts given by the COMBO-17 survey \citep{Wolf04}.
The cut in $R$ magnitude approximates the detection limit of the spectroscopic survey in the cluster, $I_{814}=23$.
Altogether, we select 62 galaxies in an area of 775 arcmin$^{2}$ ($\sim$ 20 times larger than the cluster field) and 39 of them have $f_{24} > 50 ~\mu$Jy.
Two of those field IR galaxies are classified as QSOs.
So, the fraction of the star forming galaxies in the field sample at the same 24 \micron\ threshold and of a similar redshift range is about $60\pm12$\%, much higher than the fraction we found in the cluster.
Although the photometric redshifts we used to select the field sample are less accurate than spectroscopic redshifts, the fraction of the star forming galaxies in the field does not change significantly in a large redshift range ($0.7<z_{ph}<0.95$) and therefore redshift uncertainties will have little effect on the comparison.
The smaller fraction of IR bright galaxies in the cluster compared with the field is consistent with the results given by studying the galaxy emission lines \citep{Lewis02,Gomez03} and suggest that galaxies in the cluster have a lower level of star formation on average.

Even though the cluster galaxies have a lower level of SF on average compared with the field, a fraction of 13$\pm3$\% of star forming galaxies is still very substantial considering the short IR bright phase, especially for the 6 LIRGs which constitute 4$\pm2$\% of all the spectroscopically selected cluster members.
\citet{Marcillac06a} analyzed Balmer absorption lines and the 4000${\rm \AA}$ break of a sample of LIRGs at $z\sim 0.7$ and found that the duration of the LIRG phase is most likely $\sim 0.1^{+0.16}_{-0.06}$ Gyr.
The timescale of star formation in local IR bright galaxies is even smaller, $\sim 10^7$ yr \citep{Gao04}.
If these active galaxies are due solely to infall from the field, such a short timescale would mean the cluster would have to accrete about $60^{+90}_{-37}$ LIRGs from the field per Gyr to sustain the observed SF level.
This is about half of the current spectroscopically selected cluster sample.  
Even if we consider that about 30\% of LIRGs are experiencing their second star bursts in 1 Gyr \citep{Marcillac06a}, it would still mean more than half of the cluster members are the LIRGs accreted from the field in the last Gyr.
Such a large accretion rate is very unlikely.
The average smooth growth of the cluster masses from the simulation of \citet{Rowley04} in the one Gyr period ($z\approx 1.1 - 0.83$) ranges from 10\% to 40\%.

However, it is possible that we are seeing a large fraction of star forming galaxies in this cluster due to a temporary rise of the accretion rate caused by a major merger/infall event.
The quadrupole-like temperature structure and the lack of shock-heated regions between the two X-ray peaks of this cluster suggest that the major clumps (the central and western clumps) are probably at a postmerger stage and the lack of an X-ray peak in the eastern clump may also suggest a recent infalling/passing-by \citep{Jee05a}.
These merger and infalling events might have introduced a large number of field galaxies into the cluster in a short time and boosted the accretion rate temporarily.
However, the spatial distribution of the IR galaxies does not seem to support this scenario.
Although MS 1054-03 has several subclumps and one of them has relatively enhanced SF, most of the IR galaxies do not concentrate in subclumps; instead, they tend to scatter around the cluster and avoid the two major clumps (see \S. 4.4). 
Similarly, the cluster members with 15 \micron\ emissions in Cl 0024, a mid-redshift cluster with high star formation level, also do not show spatial concentrations \citep{Coia05a}.
Such a spatial distribution does not support a major merger/infalling event.
In addition, RXJ0152 also has a similar fraction of IR bright galaxies, and they show no sign of concentration into a subclump \citep{Marcillac07}. 
This evidence suggests that such high fractions of IR galaxies in high redshift clusters might be quite common, and it is unlikely that they are all due to major infall events.
More likely, these IR galaxies have been in the cluster for quite a while.
However, they probably have never been close to the high density region before and still retain a large amount of gas.
The recent SF in these galaxies may be triggered either by interaction with the cluster intergalactic medium (IGM), with other galaxies, or by tides.
In support of this hypothesis, evidence has been found previously for star forming bursts in infalling galaxies into clusters by, e.g., \citet{Gavazzi03}, \citet{Cortese06} and \citet{Mercurio04}.

An alternative possibility is that the LIRGs in the cluster have lasted much longer than the time scale estimated by \citet{Marcillac06a}.
If the timescale is an order of magnitude longer ($\sim$ 1 Gyr), then the accretion rate would be 10 times smaller and would not raise the problem of too rapid growth.
However, such a long timescale would indicate the accreted galaxies could retain their gas for a long period and keep their star formation untouched by the cluster environment.
This alternative view is again inconsistent with the passive scenario that the star formation of field galaxies is quenched quickly after they are accreted into a cluster. 
   
\subsection{Spatial Distribution of the IR Galaxies}
\begin{figure}
\epsscale{1.0}
\plotone{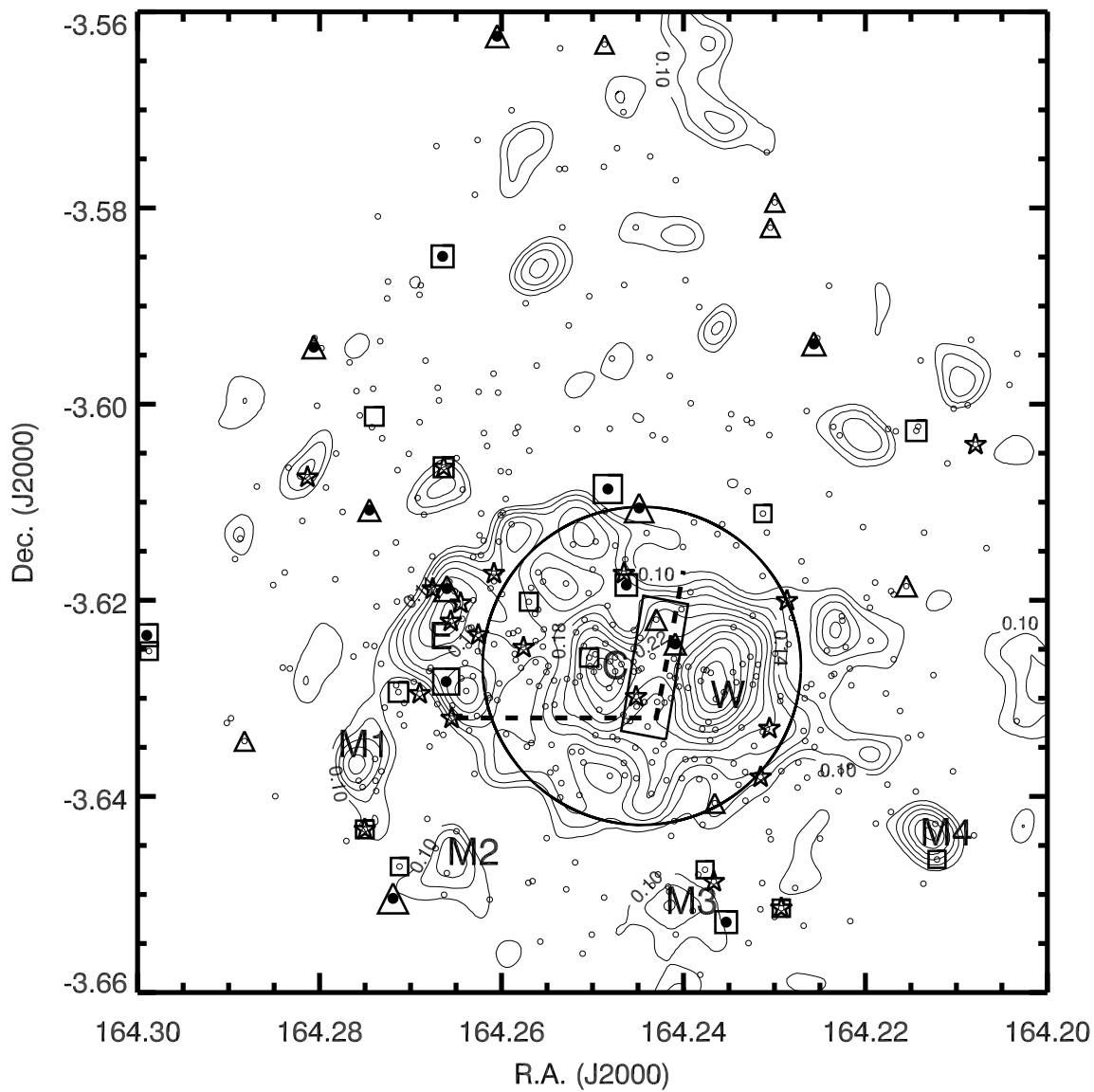}
\caption{The spatial distribution of the IR galaxies in MS 1054-03.
Open squares are the IR galaxies from the spectroscopic sample, and open triangles are the additional IR members selected by the photometric redshifts.
The sizes of the symbols are proportional to their IR luminosities.
LIRGs are also indicated by filled circles.
Small open circles are the cluster members selected by the photometric redshifts.
E+A Galaxies selected by \citet{Tran03} are indicated as open stars. 
The contours are the mass contours from \citet{Jee05a}.
The eastern, central, western and four minor mass clumps are labeled as E, C, W, and M1-M4.
The solid circle is the region with effective gas-stripping.
The rectangle is defined as the interface region of the two major clumps.
}
\label{f_mass}
\end{figure}

The spatial distribution of the IR galaxies in the cluster may help us understand the effect of the cluster environment on the galaxy SFR. 
As indicated by the X-ray and optical light distributions, the morphology of MS 1054-03 is quite complex.
\citet{Jee05a} reconstruct a high-resolution mass map of the cluster through ACS weak-lensing analysis.
They confirm the three dominant mass clumps in the cluster previously reported by a WFPC2-based weak-lensing analysis \citep{Hoekstra00} and find some detailed substructures for the first time.
In Fig.~\ref{f_mass}, we overplot the IR galaxies on this mass contour map.  
The mass map is constructed in units of the dimensionless mass density $\kappa$, and $\kappa > 0.1$ corresponds to a significance of $\gtrsim 3 $ sigma. 
The squares are from the spectroscopic sample, and the triangles are those additional members selected by photometric redshifts.
The sizes of the symbols are proportional to the IR luminosities. 
For clarity, the LIRGs are also indicated by black dots.
The three major clumps, eastern, central and western (E, C \& W), as well as the four minor clumps (M1-M4), are labeled on the plot following \citet{Jee05a}.
One distinct feature of Fig.~\ref{f_mass} is that many IR galaxies are distributed in the outskirt region of the cluster.
Two thirds of the IR galaxies are located in the low density region with $\kappa < 0.1$, and this ratio could be higher if we take into account projection effects.

Another interesting feature of the distribution is the lack of IR galaxies in the western clump and the southern extension compared to the rest of the major structure.
We divide the major body of the cluster into two approximately equal parts by the dashed line in Fig.~\ref{f_mass}, the northeastern (NE) and southwestern (SW) regions. 
The NE region includes both the eastern and central clumps and the SW region includes the western clump and its south extension.
The SW region only contains 2 or 3 IR galaxies, while the NE region contains at least 9 IR galaxies.
The ratio of the IR galaxies to the number of cluster candidates selected by photo-z in the SW and NE region is $2\pm1\%$ (3 vs. 125) and $7\pm2\%$ (9 vs. 130) respectively, compared to $10\pm2\%$ (22 vs. 213) in the outskirt region ($90\arcsec$ away from the cluster center).
Although it may arise partly from projection effects, the difference between these two regions is statistically significant.
In addition, four out of the five brightest IR galaxies of the whole cluster correlate to the NE region. 
These observations probably suggest a recent star-forming episode in the NE region.

The X-ray study of the cluster may provide some support to this scenario.
Unlike the other two major clumps, the eastern clump is absent from the X-ray map \citep{Clowe00, Jeltema01, Jee05a}.
Based on this fact and the anomalous X-ray profile of the central peak, as well as the temperature map of the region, \citet{Jee05a} proposed that the eastern clump has passed through the dense region of the central clump recently as an off-center collision, during which the intracluster gas was stripped.
It is possible that this recent off-center collision between the eastern and central clumps triggered SF in some of the IR galaxies.

By comparison, the merging between the western and central clumps is probably at a very late stage, suggested by the lack of a shock-heated region between them \citep{Jee05a}. 
We define a central $9\arcsec \times 25\arcsec$ rectangle as the interface of these two major clumps (solid box in Fig.~\ref{f_mass}); we found two out of 21 photometrically selected cluster candidates with detectable IR emission in this region, a fraction of $10\pm7\%$, comparable to the fraction in the outskirt region of the cluster.
However, we found no spectroscopically confirmed IR member in this region. 
The uncertainties of the photometric cluster candidates and the projection effects further weaken the evidence for activity at this interface. 
As a comparison, \citet{Marcillac07} found no evidence of merging triggered SF activities in the interface of the two main merging clumps in the cluster RXJ0152 at a similar redshift.
Combining the two results, clump interfaces do not appear in general to be sites of strongly enhanced star-forming activity.

\citet{Tran03} studied E+A galaxies, characterized by strong Balmer absorption and little or no [O II] $\lambda$3727 emission, in this cluster. 
The strong Balmer absorption is evidence of recent substantial star forming activities ($\leq$ 2 Gyr) and the lack of [O II] emission indicates that those galaxies probably have no current SF, and therefore they are usually classified as post-starburst galaxies \citep{Dressler83}. 
There are 19 E+A cluster members in the region covered by IR observations and they are shown as the open stars in Fig.~\ref{f_mass}.
Most of these galaxies have no IR emission and they are real post-starburst galaxies in which SF has ceased at least a few Myr ago \citep{Poggianti00}. 
One third of the post-starburst galaxies are concentrated in the NE region of the cluster, three or four are related to the SW region, and the rest are scattered outside of the main body of the cluster.
For the few post-starburst galaxies related with the SW region, only one is located in the very inner region, and the rest are distributed along the boundary region.  
This result suggests the NE region of the cluster has been active in SF for many hundreds of Myr, and the SW region, on the contrary, has been quiescent for a long time.
It is consistent with the scenario suggested by X-ray analysis, that the central and eastern clumps experienced a recent collision, while the merging between western and central clumps is at a late stage.

There are three E+A galaxies with IR emission, indicating ongoing SF. 
They are not post-starburst but rather dust-enshrouded starburst galaxies in which young stars are heavily obscured and their emission lines are extincted.
The strong Balmer absorption suggests there are also a large number of A type stars, that have probably already moved out of the dusty star-forming regions, which indicates the SF in these galaxies started at least a few Myr ago \citep{Poggianti00}.
The dusty star-forming galaxies are all distributed outside of the main body of the cluster where gas stripping is not effective and the gas fuel for SF in the galaxies is not strongly depleted.
This distribution helps explain why the SF in these galaxies continues for so long.

\subsection{Mergers and Morphologically Irregular Members}
One remarkable feature of MS 1054-03 is its high fraction of merging galaxies.
Van Dokkum et al. (1999, 2000) reported 13 ongoing mergers in this cluster, comprising 17\% of the cluster population with $L>L_{*}$.
They also classified one more galaxy (H1532) as a merging candidate but did not include it in their merger list due to the lack of spectroscopic data for its companion.
The photometric redshift suggests its companion is also a cluster member, and therefore we include it as a merger; this brings us to a total of 14 merging systems.
In addition, \citet{Tran05} found 10 bound red galaxy pairs (5 of them already in van Dokkum's sample) with projected distance smaller than 30 $h^{-1}$ kpc and relative velocity $\delta v \leq 300~ {\rm km~ s}^{-1}$.
Most of the mergers between field galaxies are accompanied by triggered SF \citep[e.g.,][]{Liu95,Patton05}.
However, \citet{Tran05} found that most of the merging galaxies in MS 1054-03 have no detectable [\ion{O}{2}] emission lines and have probably lost their gas long ago.

As pointed out in the previous section, [\ion{O}{2}] emission may underestimate the SFR in galaxies due to dust extinction and is a less robust indicator of SFR than the IR luminosity.
Therefore, we discuss the IR properties of these merging galaxies.
Among 14 merging systems, 4 have 24 \micron\ emission $f_{24} > 50~ \mu$Jy, H4683.4741, H6567, H2710 and H1532 (the last merger is the one confirmed by the photometric redshift).
The brightest one is a double nucleus, highly disturbed disk system.
The other three are interacting pairs.
One galaxy (H4822) common to two red pairs found by \citet{Tran05} has $f_{24} > 50~ \mu$Jy. 
There is another galaxy in those red pairs having weak 24 \micron\ emission ($f_{24}\approx 40 ~\mu$Jy, not included in our spectroscopic sample).
Altogether, about $29\pm16$\% of the merging systems show detectable IR emission.
For red pairs, this fraction is even lower, only about 10-20\%.

\begin{figure}
\epsscale{1.2}
\plotone{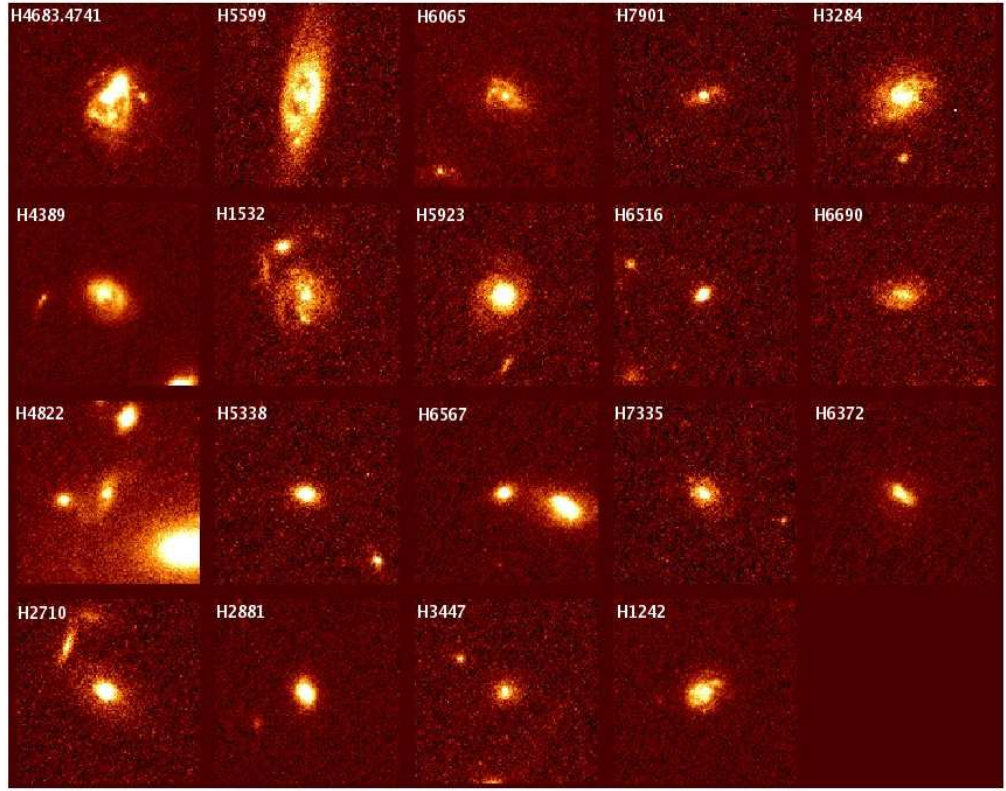}
\caption{The ACS images of spectroscopically confirmed IR galaxies in MS 1054-03. The size of each image is about $6\arcsec \times 6\arcsec$.}
\label{f_stamp}
\end{figure}

\citet{Postman05} classified the morphological types of spectroscopically confirmed cluster members of MS 1054-03.
We correlate the IR galaxies in the spectroscopic sample with their classification.
Fig.~\ref{f_stamp} shows the ACS images of these galaxies.
Without distinguishing if galaxies are in merging systems, we found $21\pm12$\% (4), 63$\pm23$\% (12) and 16$\pm10$\% (3) of the IR galaxies are early, late and irregular type galaxies, respectively.
These fractions are very similar to what \citet{Marcillac07} found for the IR galaxies in RXJ0152.
If we consider all the members of MS 1054-03 with morphological classification, 4$\pm2$\%, 52$\pm19$\% and 50$\pm35$\% of early, late and irregular type cluster galaxies have detectable IR emission ($f_{24}> 50 ~\mu$Jy).
Again, these results agree with those of \citet{Marcillac07} within the statistical errors.
Therefore, only a very small fraction of early type galaxies in the spectroscopic sample have strong SF, but half of late type and irregular galaxies have SFRs $>10~ M_{\sun}$ yr$^{-1}$.

In addition to those galaxies in merging systems, galaxies with irregular morphologies may also be considered as experiencing interactions.
Therefore, if we count the irregular galaxy (H4389) as a merger too, 6 IR galaxies out of 19 (32$\pm15$\%) are related to galaxy merging/interaction.
Furthermore, two IR galaxies (H6065 and H6372) classified as normal late type galaxies also show some irregular features.
If their irregular features are also related to galaxy merging/interaction, then 42$\pm18$\% of the IR galaxies may have galaxy merging/interaction triggered SFs.

\section{Discussion}
\subsection{Comparison with RXJ0152.7-1357} 
From \S 4.2, we found that the IR LF of MS 1054-03 evolves as strongly as the field IR LF and has an over abundance of IR galaxies down to log$L_{IR}= 44.6$, comparable to the field IR LF at the same redshift.
This result is quite surprising given studies showing decreased SF in rich clusters.
However, with only one cluster, it is not clear if the result is typical of rich clusters at $z\approx0.8$ or peculiar to MS 1054-03.

The study of IR galaxies in RXJ0152 by \citet{Marcillac07} provides a good comparison.
The 24 \micron\ data for these two clusters have similar sensitivities and the optical data cover both central $5\arcmin \times 5\arcmin$ regions of the clusters.
X-ray data show that RXJ0152 also has two regions with peak emission \citep[e.g.,][]{Huo04}, and the dynamical analysis of the cluster suggests an ongoing merger in the system \citep{Girardi05}.
In addition, photometric and spectroscopic surveys by \citet{Kodama05} and \citet{Tanaka06} discovered two large-scale filament-like structures hosting the central main cluster.
\citet{Marcillac07} found 22 IR galaxies in RXJ0152 confirmed by spectroscopic data as cluster members. 
We plot the IR LF deduced from their data as the open stars in Fig.~\ref{f_LF}.
We did not make any incompleteness correction for the IR LF of RXJ0152 because their sample only includes galaxies with $f_{24}$ above the 80\% completeness limit and the spectroscopic data in this cluster are quite deep, complete down to $R=24$ \citep{Demarco05}. 
Therefore, the IR LF of RXJ0152 without any correction is comparable to that of MS 1054-03 with incompleteness corrections above the 80\% completeness limit at 24 \micron.

As stated in \S 3.5, \citet{Marcillac07} use a different set of SEDs to deduce a total IR luminosity from the 24 \micron\ flux.
Although it does not result in a significant difference, it is better if we compare the IR LF of RXJ0152 with that of MS 1054-03 deduced from the same method (open and filled triangles in Fig.~\ref{f_LF}).
There is a large difference in the faintest data points of the IR LF of RXJ0152 and MS 1054-03, due to the different cutoffs of 24 \micron\ flux density in the two samples: the RXJ0152 sample is only determined down to the 80\% limit ($f_{24} > 83~ \mu$Jy) while the MS 1054-03 sample, after incompleteness correction, is derived down to the 50\% limit ($f_{24} > 50~ \mu$Jy).
The dashed vertical line in Fig.~\ref{f_LF} shows the IR luminosity corresponding to the 80\% limit.
The lowest bin is partly below this limit.
For both of the two brighter data points, RXJ0152 has a higher value than MS 1054.
However, the differences are still within one sigma Poisson errors.
The brightest data point for RXJ0152 includes 5 IR galaxies, while the data point for MS 1054-03 only includes 2 IR galaxies from the spectroscopic sample.
The photometric selection adds two more galaxies in MS 1054-03 within this luminosity range, but as discussed before, those two are probably background sources.
The IR LF of RXJ0152 confirms the strong evolution trend and the over abundance of bright IR galaxies we found in MS 1054-03 and it shows an even larger number of the brightest IR galaxies as pointed out by \citet{Marcillac07}.

\citet{Marcillac07} find that most of the IR galaxies of RXJ0152 are distributed outside of the two major clumps indicated by X-ray emission.
They also find a larger median redshift for these galaxies compared to the cluster redshift, which is identical to the larger median redshift of the infalling late-type cluster members found by \citet{Blakeslee06}. 
Based on these facts, they suggest that infall of galaxies is probably responsible for much of the star formation activity we see in the cluster.
In MS 1054-03, we did not find a difference in the redshifts of the IR galaxies compared with the rest of the cluster members.
However, about 60\% of the IR galaxies are located outside of the main body of the cluster (major clumps E, C and W).
The projection effect may make the fraction even larger.
Some of these galaxies probably correspond to the infalling galaxies found by \citet{Marcillac07}.
As discussed in \S 4.3 and \S 4.4, it is unlikely that the cluster only passively accretes star forming galaxies from the surrounding field and those galaxies have a high level of SF simply due to their recent origination from the field.
On the contrary, it is very possible that we are seeing an increased SFR in infalling galaxies triggered by the galaxy-IGM interaction as shown by previous theoretical and observational evidence \citep{Fujita99,Gavazzi95,Gavazzi01}.

However, we can not rule out other mechanisms being responsible for the SF in these galaxies.
In MS 1054-03, about one third of the outside IR galaxies are associated with some minor clumps of the cluster.
Their SF may arise from processes more common in the group environment, e.g., galaxy interactions \citep{Lewis02}. 
For the 40\% of the IR galaxies correlated with the main body of the cluster, the majority are associated with the NE region of the cluster where a collision of subclumps might have occurred recently.
This result indicates the interaction of the subclumps and the processes accompanied with it, e.g., the tidal gravitational field, may also play a role in triggering SF and cause the concentration of IR galaxies \citep{Bekki99}.

The difference in the mass of the two clusters can complicate the comparison.
Using the same weak-lensing technique and data of similar quality, \citet{Jee05a, Jee05b} produced enclosed mass profiles for both of them.
The profiles show that MS 1054-03 is much richer and more massive than RXJ0152. 
The enclosed mass within 1 Mpc of MS 1054-03 is about two times as large as that in RXJ0152.
Therefore, with the similar $5\arcmin \times 5\arcmin$ IR and spectroscopic/photometric coverage of the clusters, we actually only observe the inner part of MS 1054-03 but reach the outside region in RXJ0152, where most of its LIRGs reside.
It is possible that we would find more infalling IR galaxies if our IR and spectroscopic/photometric data extended further to the outer regions of MS 1054-03.
The many IR galaxies distributed at the very edge of the survey region (see Fig.~\ref{f_mass}) seem to support this argument.
This would also explain the slightly higher IR LF of RXJ0152 compared with MS 1054-03.

Although both MS 1054-03 and RXJ0152 have two X-ray peaks and two corresponding major clumps indicating a merger, MS 1054-03 lacks shock-heated regions between the two X-ray peaks \citep{Jee05a}, but RXJ0152 has excess X-ray emission between the two clumps suggestive of a shock front \citep{Maughan03}.
The differences may indicate the different merger stage the two clusters are in: MS 1054-03 is probably at a post-merger stage while RXJ0152 is at a pre-merger stage.
This difference may contribute to the slightly larger star formation rate in RXJ0152 than in MS 1054-03.

\subsection{Evolution of the Integrated SFR in Clusters}
We have already compared the IR LF of MS 1054-03 to the Coma cluster and found a strong evolution in both $\phi^{*}$ and $L_{IR}^{*}$.
Another way to compare the SFR in clusters is to compare their integrated SFRs within a certain radius.
Using the SFR measured from emission lines, \citet{Finn04,Finn05}, \citet{Kodama04}, and \citet{Homeier05} compared the integrated SFRs of several clusters within 0.5$R_{200}$.
They also compared the integrated SFRs normalized by the cluster masses.
The mass-normalized integrated SFRs are comparable to the fractions of star-forming galaxies in the clusters, which are widely used in many systematic studies of cluster SF.
Some suggestive correlations between the integrated SFRs and redshifts/masses of the clusters were found.
However, the results are very uncertain.

There are concerns about the usual methods for estimating the cluster masses.
\citet{Finn05} used velocity dispersion, while \citet{Homeier05} suggest X-ray temperature may be a better indicator of the mass.
However, velocity dispersion and X-ray temperature are valid mass estimators only in relaxed clusters under the assumption of hydrodynamic equilibrium, which is often questionable, especially for clusters at high redshift. 
To clarify the results found in those studies, we add more data points by including the integrated SFRs of three clusters observed by MIPS (Coma, MS 1054-03, RXJ0152), and four clusters observed by ISOCAM (A2218, A1689, A2219, Cl 0024).
The seven clusters with the SFRs measured from $H{\alpha}$ emission are also added (A1367, AC114, A2390, Cl 0023, Cl 1040, Cl 1054 and Cl 1216).
Despite the systematic difference between the SFR measured from the emission line strengths and from the IR luminosity, there is general agreement after extinction correction.
We also add an average value for clusters at $0.3<z<0.5$ deduced from $IRAS$ data.
For the cluster mass, we take the mass measured from lensing analysis whenever it is available because it is free from any assumptions about the dynamical state of the clusters. 
We limit our calculation to within the 0.5$R_{200}$ region and set the cutoff in the SFR as 2 $M_{\sun}$ yr$^{-1}$.
The references and details about the integrated SFR and the masses of these clusters are provided in Appendix A. 

In panels $a$ and $b$ of Fig.\ref{f_intsfr}, we plot the integrated SFRs as a function of redshift and cluster mass. 
In panels $c$ and $d$, we plot the mass-normalized values.
The integrated SFRs show a weak evolution with redshift.
The evolution is more pronounced in the mass-normalized SFRs, approximately $\propto (1+z)^5$.
The nonparametric Spearman tests show that the significances of the correlations are 97\% and 99\% for the integrated SFRs and the mass-normalized integrated SFRs.  
However, this evolution trend is complicated by the anticorrelation between the mass-normalized integrated SFRs and the cluster masses.
Although the integrated SFR does not show an apparent correlation with mass, the mass-normalized one has an anticorrelation with mass, $\propto M^{-0.9}$, with a Spearman significance of 97\%.
This anticorrelation agrees with the results found in previous comparisons \citep{Homeier05,Finn05}.
It also agrees with the anticorrelation found by \citet{Poggianti06} between the fraction of star-forming galaxies and the cluster velocity dispersion, in the sense that the fraction of star-forming galaxies is comparable to the mass-normalized integrated SFR.
Given this anticorrelation, the evolution we found in the mass-normalized integrated SFR is probably due to the different masses of the low-redshift and high-redshift clusters in our sample.

As shown in panel $e$ of Fig.\ref{f_intsfr}, the clusters with $z<0.5$ in our sample are on average more massive than those with $z>0.5$.
Therefore, the increased mass-normalized integrated SFRs at higher redshifts found in the sample could merely be a selection effect.
To disentangle the mass factor from the evolution trend, we select a subsample of clusters with a mass range of $3\times10^{14}~M_{\sun}<M<12\times10^{14}~M_{\sun}$, in which both low-redshift and high-redshift clusters have good sampling, and plot their mass-normalized SFRs vs. redshift in panel $f$.
The evolution trend of the subsample becomes much weaker, but it still has a significance of about 89\%. 
In addition, even if the evolution of the mass-normalized integrated SFR can be largely caused by the anticorrelation between the mass-normalized integrated SFR and the cluster mass, the evolution of the integrated SFR without mass normalization is still significant and can not be easily explained by the mass selection effect. 

Among all the clusters in our sample, Cl~0024 has the largest integrated SFR with an intermediate redshift and cluster mass.
Its integrated SFR is at least five times larger than those clusters with similar masses.
Using MIPS 24 \micron\ data, \citet{Geach06} also found a very significant excess of mid-infrared sources up to $r<5$ Mpc in Cl~0024 compared to another cluster MS 0451-03 at $z=0.55$. 
These results suggest that Cl~0024 is quite unusual compared to other clusters in the sample.
Another unusual aspect of Cl~0024 is its relatively faint X-ray emission compared to its large mass.
It has a mass a little larger than RXJ0152, but its X-ray luminosity is only about a fifth of RXJ0152.

The eastern clump of MS 1054-03 also has significant star formation activities but is absent in X-ray emission, contradicting the expected $L_{X}$ from its mass \citep{Jee05a}.
For a comparison, we calculate the integrated SFRs for the eastern clump of MS 1054-03 separately and plot it as an open star in Fig.~\ref{f_intsfr}.
This result has the interesting implication that clusters or cluster subclumps with unusually low X-ray emission may have very active SF.
A recent work by \citet{Popesso07} seems to support this conclusion.
\citet{Popesso07} studied 137 Abell clusters and found that clusters with lower X-ray luminosity than expected from the $L_{X}-M$ relation, the so-called X-ray-Underluminous Abell clusters (AXU), show a velocity distribution characteristic of accretion and have a higher fraction of blue galaxies in their outer regions.
They suggest the low X-ray luminosities of these clusters are due to the ongoing accretion or merging process.
Although Cl~0024 is not exactly X-ray underluminous according to their definition, the exceptionally high SFRs in it and in the eastern clump of MS 1054-03 generally agree with the scenario they propose.
They also found that about 40\% of the clusters they studied are AXU, indicating that AXU clusters are not a small minority, at least at $z<0.4$, and suggesting they probably host more SF in total than the X-ray-luminous clusters.
Since most of the clusters in our sample are X-ray luminous ones, the SFRs have probably been biased towards the lower value and have an evolution reflecting only conditions in well relaxed systems with substantial amounts of hot, X-ray emitting plasma.

As pointed out previously, the star-forming galaxies tend to be located in the outer regions beyond 0.5$R_{200}$.
Thus, the integrated SFRs within 0.5$R_{200}$ only present a portion of the total star forming activities in the cluster.
This cutoff effect is especially significant for RXJ0152 due to its irregular morphology.
The integrated SFR within 0.5$R_{200}$ of RXJ0152 only accounts for about 13$\pm3$\% of its total SFR in the survey region, much smaller than the 50\% expected from a singular isothermal (SIS) distribution of the star-forming galaxies.  
For MS 1054-03, this fraction is higher, 70\%, comparable with 66\% expected from a SIS distribution.

\subsection{Ram Pressure Stripping}
In the preceding section, we showed that the mass-normalized integrated SFRs of clusters have an anticorrelation with the masses.
This result indicates that massive clusters are probably more effective in suppressing SF than the low mass ones \citep[see also,][]{Poggianti06}. 
We also found that very few IR galaxies in MS 1054-03 are distributed in the region of high mass density, especially in the southwestern part.
On the other hand, gas stripping by ram pressure in clusters \citep{Gunn72} is also found to be more pronounced in the massive clusters and more effective in the high density regions \citep[e.g.,][]{Giovanelli85}.
The coincidence may suggest SFR suppression in the clusters due to gas stripping by the ram pressure of the intracluster medium (ICM).

To investigate the effect of the ram pressure gas stripping, following \citet{Homeier05}, we calculate the effective radius for this process.
\citet{Jee05a} found the X-ray surface brightness profile of the cluster is best fitted by an isothermal $\beta$ gas model with $\beta=0.78\pm 0.08$ and $r_{c}=16\arcsec\pm 15\arcsec$.
Due to the complex morphology of the cluster, this isothermal form does not fit the surface brightness profile of the inner $r<45\arcsec$ region, but it fits the outer region very well and predicts a projected mass profile consistent with the result from the weak-lensing analysis.
The virial radius of the cluster is $1.7 \pm 0.2$ Mpc and the corresponding virial mass is $1.2 \pm 0.2  \times 10^{15}$ M$_{\sun}$ \citep{Jee05a}.
If we assume the gas mass is about 10\% of the total mass, as suggested in \citet{Neumann00}, we can deduce a central gas density $\rho_{0,gas}=8 \times 10^{14} $ M$_{\sun}$ Mpc$^{-3}$.
According to the ram pressure stripping criterion for the gas in a disk \citep{Fujita99} and the gas profile of MS 1054-03, we obtain a ram pressure stripping effective radius of $r_{rp} = 63 \arcsec$ ($\sim$ 0.5 Mpc) for a Milky Way-type galaxy with a velocity of 1000 km s$^{-1}$.

We show the $r_{rp}$ in Fig.~\ref{f_mass}.
For RXJ0152, $r_{rp}$ $\sim$ 0.3 Mpc \citep{Homeier05} and none of its star-forming galaxies lies within this radius (also see the discussion in Marcillac \etl, 2007).
In MS 1054-03, the majority (80\%) of the IR cluster galaxies are distributed outside of $r_{rp}$, providing strong evidence of SF suppression due to ram pressure stripping.
For the six IR galaxies within $r_{rp}$, only two of them are brighter than $f_{24} = 80~ \mu$Jy.
The fraction of IR galaxies ($4\pm2\%$) in this region is about half of that in the outskirt region but still larger than the fraction in the southwestern part, which supports the suppression effect of ram pressure stripping, but also suggests it is less effective for some galaxies and $r_{rp}$ is only an approximate measurement of its effectiveness.
There may be a number of explanations.
First, this could just be a projection effect, that is at least some of the six galaxies lie in front of or behind the ram stripping region.
Second, the isothermal density profile is only an approximate description of the gas distribution, so $r_{rp}$ is only a rough indicator of the effectiveness of ram pressure stripping.
Third, the ram pressure stripping criterion we used to deduce $r_{rp}$ is only for the gas in the disk of a galaxy.
The gas in the inner disk of a galaxy is harder to strip and therefore any SF occurring there, e.g., circumnuclear SF, is more difficult to suppress.

Another interesting fact is that about half of the E+A galaxies are distributed along the effective radius.
This behavior suggests gas stripping as the reason that the star forming activity stopped in these galaxies.

\subsection{Galaxy Interaction}
Many studies find that most of the star-forming galaxies in both the field and clusters are morphologically not strongly disturbed at $z\sim0.8$ \citep[e.g.,][]{Bell05,Marcillac07} and argue that mechanisms that would dramatically disturb the morphology of a galaxy, e.g. strong galaxy interactions, can not be the major factor causing the change of SFRs with epoch and environment.
However, in MS 1054-03, we found 5 out of 19 IR galaxies of our spectroscopic sample are in interacting systems, one isolated galaxy is irregular, and two more have some irregular features.
Therefore, more than 30\% of the star-forming galaxies in the spectroscopic sample may be related to galaxy interactions.
Among 5 IR galaxies in interacting systems, only two have disturbed morphology, and the other three look regular.
If we only count the irregular galaxies as mergers, no matter if they are in merging systems or not, only 16\% (3 out 19) of the IR galaxies are related to galaxy interaction, consistent with the result from \citet{Bell05}, who found less than 30\% of field IR galaxies are strongly interacting at $z\sim0.7$.

For the regular IR galaxies in interacting systems, it seems that the galaxy interactions may have triggered their star formation activity but do not change their morphology sufficiently to be obvious.
This could be due to the different time scales on which SF and morphological change occur during a galaxy interaction.
It is also possible that those interactions are only strong enough to trigger the instability of the galaxies and cause star formation activity, but not to cause observable morphology distortions.
Interaction triggered SF is not always accompanied by disturbed morphology and the study of galaxy morphology at the redshift of MS 1054-03 is probably only sensitive to the strongly interacting systems.

On the other hand, the majority of the bright interacting systems ($\sim$ 70\%) in the cluster do not have detectable IR emission.
Some of them have strong interacting features, e.g., double nuclei, distorted morphologies, but these characteristics are not accompanied by strong SF.
This is probably because most of the interacting galaxies have already lost their gas (dry merger) while falling into the cluster due to, e.g., ram pressure stripping, and cannot support a high level of SF.
An example of a dry merger with little star formation increase has also been found among local galaxies \citep[e.g.,][]{Boselli05}.

\section{Conclusion}
Using the MIPS 24 \micron\ data for the rich cluster MS 1054-03 at $z=0.83$, we found 19 IR emitting cluster members selected by spectroscopic data and 15 additional IR cluster member candidates selected by photometric data.

We constructed the IR luminosity function of the cluster and find a strong evolution when compared with the IR LF of the Coma cluster with a similar mass at $z=0.02$.
The characteristic IR luminosity ($L_{IR}^{*}$) of MS 1054-03 is about one order of magnitude larger than that of the Coma cluster.
The SFR density integrated from the IR LF is about 16 times larger than that in the Coma cluster.
The evolutionary trend of the IR LFs from Coma to MS 1054-03 is similar to the evolution of the IR LFs in the field.
The comparison of the mass normalized integrated SFR of MS 1054-03 with several other clusters seems to agree with the evolution suggested by the IR LFs, but it is less conclusive because of the combined mass and redshift dependence of the SFR.
The similar SFR evolution in the clusters and in the field favors some internal mechanism, e.g., the consumption of the gas fuel in galaxies, as being responsible for the decline of SFR in different environments.

A substantial fraction (13$\pm3$\%) of cluster galaxies are forming stars actively.
Although the fraction is lower than that in the field (52\%), the overdensity of the IR galaxies in the cluster is still quite high, $\sim$20.
Such a high level of SF is evidence against the scenario that the cluster is only accreting star-forming galaxies from the surrounding field passively, after which their star formation is quenched.
Instead, it appears that many cluster galaxies continue to form stars at a high rate.
A number of cluster galaxies still have large amounts of gas and their SF can be triggered by the interactions with the intergalactic medium, with other galaxies, or, by tides.
However, there are few IR galaxies distributed in the high density regions of the cluster, indicating the suppression effect of ram pressure stripping on the SFR in those regions.
Both the IR galaxies and the E+A galaxies of the cluster show a concentration in the NE region of the cluster, supporting the scenario that an interaction between subclumps occurred recently and enhanced the SFR. 

About half of the bright late type and irregular cluster galaxies have detectable IR emission, but for early type galaxies this fraction is only about 4\%.
Only 29\% of the mergers in the cluster have detectable IR emission. 
The majority of the mergers probably have lost their gas fuel long ago and can not support a high level of SF. 
More than 30\% of IR galaxies show evidence of galaxy interaction, and only half of them have irregular morphologies, suggesting the interaction-triggered morphological change and star formation activities of galaxies have different time scales and intensities.

\acknowledgments
This work was supported by funding for \textit{Spitzer} GTO programs by NASA,
through the Jet Propulsion Laboratory subcontracts \#1255094 and \#1256318.
We thank Marijn Franx and Stijn Wuyts from FIRES group for their help with IRAC data.
We acknowledge Pablo P\'{e}rez-Gonz\'{a}lez for providing additional photometric redshift confirmation and Emeric LeFloc'h for information of IR galaxies in the CDF-S field.  
L. Bai thank Casey Papovich for helpful discussions and Aleks Diamond-Stanic for comments on the writing.
K. Tran acknowledges support from the NSF Astronomy \& Astrophysics Postdoctoral fellowship under award AST-0502156 and from the NOVA fellowship program.

\appendix
\section{Integrated SFRs and Masses}
The integrated SFR of MS 1054-03, RXJ0152 and Coma, are 372, 134 and 35 $M_{\sun}$ yr$^{-1}$, respectively.
Since the MIPS IR data are only complete down to $\sim 10~ M_{\sun}$ yr$^{-1}$ for MS 1054-03 and RXJ0152, we have applied a correction factor of 1.5 estimated from their best fitted IR LFs to the observed integrated SFRs.
We adopt weak-lensing masses of $1.1\pm0.1\times10^{15}~M_{\sun}$ \citep{Jee05a} and $4.5\pm2.7\times10^{14}~M_{\sun}$ \citep{Jee05b} for MS 1054-03 and RXJ0152.
For the Coma cluster, we use a mass of $1.4\pm0.4\times10^{15}~M_{\sun}$ from the dynamical analysis \citep{Lokas03}.

The four clusters observed by ISOCAM are A2218 \citep{Biviano04}, A1689 \citep{Duc02}, A2219 \citep{Coia05b} and Cl 0024 \citep{Coia05a}.
We adopt their masses as $4.8\pm1.4\times10^{14}~M_{\sun}$ \citep{Pratt05}, $1.93\pm0.2\times10^{15}~M_{\sun}$ \citep{Broadhurst05}, $1.0\pm0.7\times10^{15}~M_{\sun}$ \citep{Dahle06} and $5.7\pm1.1\times10^{14}~M_{\sun}$ \citep{Kneib03}.
These are all lensing masses, except for A2218, which is measured by fitting the X-ray temperature profile.
The ISOCAM surveys of A2218, A1689 and A2219 only covered the central $\sim 0.2R_{200}$ regions.
Following \citet{Finn04}, we correct for the small coverage by assuming the star-forming galaxy distribution has a singular isothermal (SIS) profile.
This correction will give a lower limit to the integrated SFR if the star-forming galaxies are more likely to reside in the outskirt region of the cluster as suggested by MS 1054-03 and RXJ0152.
For A2219 and Cl 0024, we also need to correct for the incompleteness of the detection limits, which are both $\sim 10~M_{\sun}$ yr$^{-1}$.
Since we do not have IR LFs for these two clusters, we use the IR LFs of field galaxies at similar redshifts to estimate the correction to extend the detection limit down to $2~M_{\sun}$ yr$^{-1}$.
This correction is made on the assumption that the shape of the IR LF of the rich cluster does not differ significantly from that of field galaxies at the same redshift, which is probably true given Coma and MS 1054-03 as examples.
After these corrections, the integrated SFRs for A2218, A1689, A2219 and Cl 0024 are: $14$, $64\pm17$, $307$ and $753~M_{\sun}$ yr$^{-1}$.
For the SFRs deduced from the IR luminosity, we assume a 50\% error if it has not been given, considering the typical uncertainties of the IR flux measurement.

We include seven clusters with SFRs measured from $H{\alpha}$ emission in the comparisons, A1367 \citep{Iglesias02}, AC114 \citep{Couch01}, A2390 \citep{Balogh00}, Cl 0023 \citep{Finn04} and three clusters from \citet{Finn05}: Cl 1040, Cl 1054, Cl 1216.
The SFRs of these clusters are measured from $H{\alpha}$ narrowband imaging, except AC114 which is measured from $H{\alpha}$ spectroscopy. 
The mass of A1367 is estimated using the dynamical analysis, $7.1\pm1.5\times10^{14}~M_{\sun}$ \citep{Girardi98}.
We use the lensing mass of $7.3^{+4.4}_{-1.9}\times10^{14}~M_{\sun}$ \citep{Natarajan98} for AC114.
For A2390, we adopt a mass of $13.6\pm0.7\times10^{14}~M_{\sun}$ from X-ray analysis \citep{Allen01}, which gives a consistent result with the lensing analysis in the inner region where it is available \citep{Squires96}.
Following \citet{Kodama04}, we use a dynamical mass of $2.3\pm1.2\times10^{14}~M_{\sun}$ for Cl 0023 \citep{Postman98}.
For Cl 1040, Cl 1054, and Cl 1216, we use the lensing masses of $0.55^{+0.75}_{-0.48}\times10^{14}~M_{\sun}$, $4.8^{+1.5}_{-1.4}\times10^{14}~M_{\sun}$ and $9.5^{+1.8}_{-1.8}\times10^{14}~M_{\sun}$ \citep{Clowe06}, respectively.
For A1367, we correct for the incomplete coverage both in space and velocity range by a factor of 1.7 \citep{Iglesias02} and obtain an integrated SFR of $29.1\pm3.1 ~M_{\sun}$ yr$^{-1}$. 
For AC114, we also include a correction factor of $~2.8$ for the sampling fraction and aperture bias of the spectroscopic survey following \citet{Kodama04}.
All of these $H{\alpha}$ surveys are complete down to $2~M_{\sun}$ yr$^{-1}$ and we make a correction for incomplete survey coverage for Cl 1216.
The integrated SFRs for AC114, A2390 and Cl 0023 are $21.6 \pm 19.5$, $80\pm28$ and $71\pm23 ~M_{\sun}$ yr$^{-1}$.
These values are slightly smaller than those given by \citet{Homeier05} because we limit the integration to galaxies with SFR $>2~M_{\sun}$ yr$^{-1}$.
For Cl 1040, Cl 1054 and Cl 1216, we obtain integrated SFRs of $63\pm12$, $90\pm19$ and $369\pm55 ~M_{\sun}$ yr$^{-1}$.
The results for Cl 1040 and Cl 1216 are similar to those found by \citet{Finn05}.
However, because the lensing mass of Cl1054 is about three times larger than the mass estimated from the velocity dispersion by \citet{Finn05}, the integrated SFR is also about three times larger than that given by \citet{Finn05} due to the larger $R_{200}$.

In addition, the SFRs of these seven clusters are deduced by assuming a H${\alpha}$ extinction of 1 mag, which corresponds to $A_{V}\sim1.2$ mag.
However, as mentioned in section 4.3, most of the star-forming galaxies have extinctions larger than this value.
For example, all the IR galaxies in MS 1054-03 have $A_{V}^{IR}$ larger than 1.5 mag.
To account for the underestimate of the extinction for the H${\alpha}$ deduced SFRs, we need to know the average $A_{V}$ of galaxies in each cluster.
Since we do not have the $A_{V}$ measured for these galaxies, we have to assume the shapes of the IR LFs of clusters are not very different from those of field galaxies at the same redshifts and estimate the average IR luminosities for those galaxies with SFR $>2~M_{\sun}$ yr$^{-1}$.
This assumption may be inaccurate, but both Coma and MS 1054-03 IR LFs seem to support it.
From the average IR luminosities, we can estimate the $A_{V}^{IR}$ with the IR-luminosity-dependent extinction formula given by \citet{Choi06}.
Comparing those $A_{V}^{IR}$ with the assumed 1 mag extinction at H${\alpha}$, we deduce correction factors for the integrated SFRs in these clusters, which range from 1.2 to 1.5.
We apply the corrections for all the H${\alpha}$ deduced integrated SFRs when comparing them with those deduced from IR luminosities.

Using $IRAS$ 60 \micron\ data, \citet{Kelly90} find the average flux of a sample of clusters by stacking and folding their addscan signals.
There are 58 clusters in their sample with $0.3<z<0.5$, and their average flux is about 29.1 mJy.
(They also calculated a value for a sample of local clusters with $z\sim 0.05$, but due to the uncertainties caused by the different techniques used to deduce the average flux, we do not consider those clusters here.) 
Each scan of the $z>0.3$ clusters is about $20\arcmin$ long and $5\arcmin$ wide and the signal is recorded as the one dimensional flux along the scan.
The width of the scan is approximately equal to $R_{200}$ of a cluster with $7\times10^{14}~M_{\sun}$ at this redshift and the PSF fitting of the scan signal along the length includes the flux approximately from the central $5\arcmin$ region.
So the average flux they obtained comes from a similar region as our integrated SFR for a cluster with $7\times10^{14}~M_{\sun}$.
For the clusters with larger/smaller mass than $7\times10^{14}~M_{\sun}$, the flux is an underestimate/overestimate of the total flux within 0.5$R_{200}$ regions due to the limitation of the scan region.
Because the evolved Coma IR LF suggests that about one third of the total IR luminosity is coming from LIRGs at this redshift, we deduce a composite $L_{IR}/L_{60}$ ratio for the clusters by averaging the ratio given by LIRG and late type SEDs weighted by this factor.
Using this composite $L_{IR}/L_{60}$ ratio, we can correlate the 60 \micron\ flux of the cluster to the total IR luminosity and therefore the total SFR.
The error of this total SFR is dominated by the uncertainties caused by the composite $L_{IR}/L_{60}$ ratio.  
Since the average 60 \micron\ flux of these clusters is obtained by stacking, it is not limited by the detection limit of the observations and the total SFR deduced from it also includes the contribution from the galaxies with SFR $<2 M_{\sun}$ yr$^{-1}$. 
Those galaxies contribute 30\% of the total SFR, estimated from the IR LF.   
Using this factor, we convert the average total SFR to the average integrated SFR.
Assuming a typical mass of $7\times10^{14}~M_{\sun}$ for these clusters, we obtain the mass-normalized integrated SFR.
The assumption of the typical mass here will cause some uncertainty on this data point, but it will partly offset the effect caused by the limitation of the scan region.

\bibliographystyle{apj}

\clearpage

\input{tab1.tex}

\clearpage

\begin{figure}
\epsscale{1.0}
\plotone{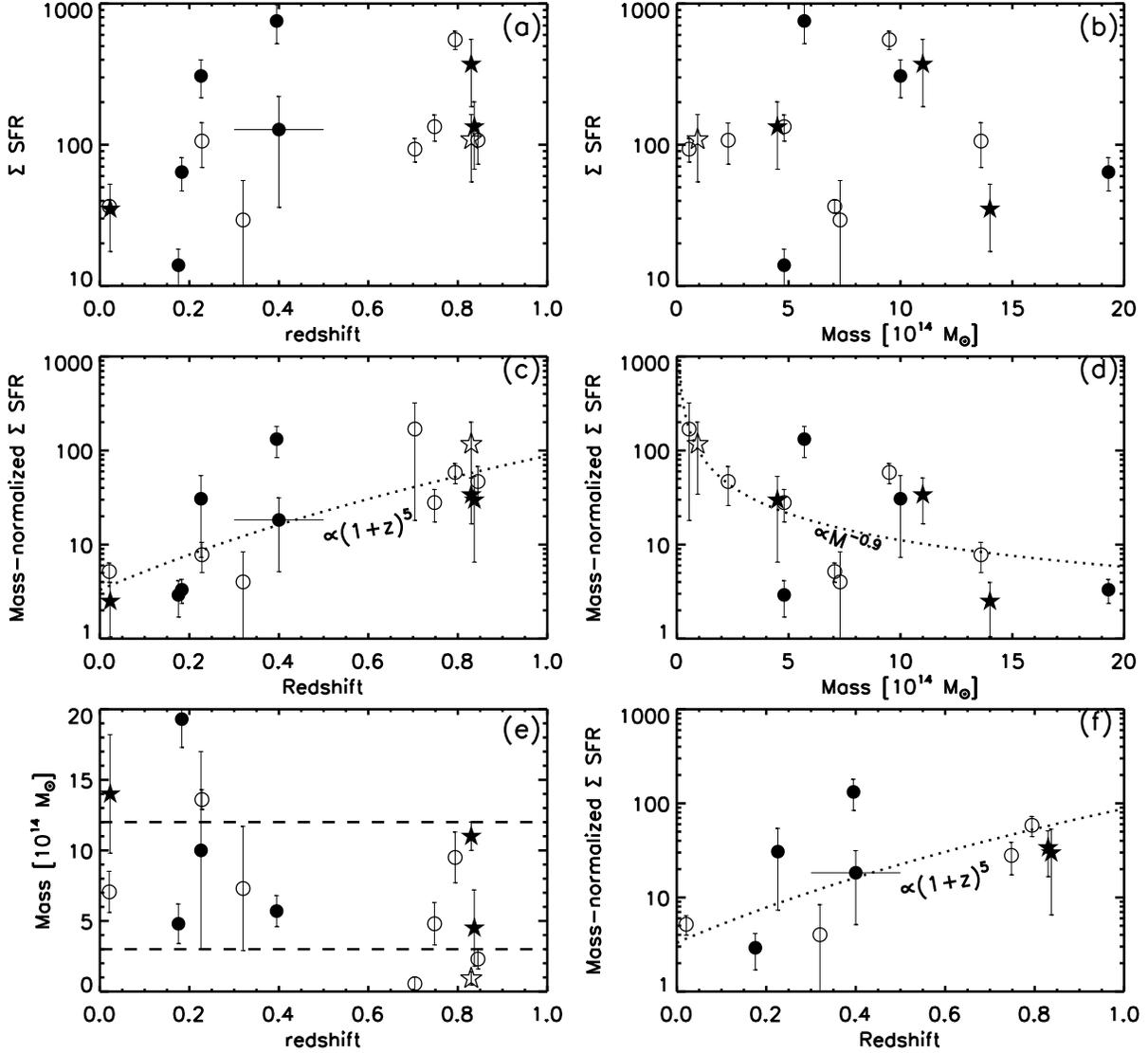}
\caption{($a$) and ($b$), the integrated SFRs vs. redshifts and cluster masses;
($c$) and ($d$), the mass-normalized SFRs vs. redshifts and cluster masses;
($e$), the cluster masses vs. redshifts;
($f$), the mass-normalized SFRs vs. redshifts for clusters with $3\times10^{14}~M_{\sun}<M<12\times10^{14}~M_{\sun}$. 
The mass selection limits are indicated as the two dashed horizontal lines in ($e$). 
The filled stars are three clusters observed with MIPS: Coma, MS 1054-03, and RXJ0152.
The open star is the eastern clump of MS 1054-03.
Filled circles are the clusters observed with ISOCAM, and open circles are from H${\alpha}$ emission line measurements.
The dotted curves in ($c$) and ($f$) are the fitted correlation between mass-normalized SFRs and redshifts for all the clusters (not including the eastern clump of MS 1054-03).
The dotted curve in ($d$) is the fitted correlation between mass-normalized SFRs and masses (not including the eastern clump of MS 1054-03).
}
\label{f_intsfr}
\end{figure} 
\end{document}

%% file: tab1.tex
\begin{deluxetable}{rrrr}
\tablecolumns{4}
\tablewidth{0pc}
\tablecaption{Cluster galaxies with 24 \micron\ emission}
\tablehead{
\colhead{ID\tablenotemark{a}}&\colhead{$K_{s,AB}$\tablenotemark{a}}&\colhead{$f_{24}$\tablenotemark{b}}&\colhead{log$L_{IR}$}\\
\colhead{}&\colhead{(Mag)}&\colhead{($\mu$Jy)}&\colhead{(ergs s$^{-1}$)}
}
\startdata
\multicolumn{4}{c}{Spectroscopic Sample}\\
\hline
1140 & 20.24$\pm$ 0.06& 363$\pm$ 33&45.37\\
 553 & 19.83$\pm$ 0.02& 317$\pm$ 30&45.13\\
1663 & 20.51$\pm$ 0.05& 140$\pm$ 28&44.69\\
 791 & 21.90$\pm$ 0.09&  76$\pm$ 44&44.68\\
 102 & 19.86$\pm$ 0.06& 184$\pm$ 32&44.68\\
 909 & 20.76$\pm$ 0.01& 116$\pm$ 24&44.63\\
1316 & 20.91$\pm$ 0.01&  96$\pm$ 26&44.52\\
1200 & 20.07$\pm$ 0.07& 107$\pm$ 36&44.45\\
 581 & 21.61$\pm$ 0.03&  62$\pm$ 30&44.43\\
1357 & 20.86$\pm$ 0.04&  82$\pm$ 25&44.32\\
 695 & 20.85$\pm$ 0.02&  76$\pm$ 22&44.29\\
 874 & 20.68$\pm$ 0.04&  68$\pm$ 28&44.25\\
 211 & 19.97$\pm$ 0.02&  63$\pm$ 31&44.22\\
 725 & 21.50$\pm$ 0.08&  51$\pm$ 34&44.21\\
 170 & 21.14$\pm$ 0.06&  59$\pm$ 25&44.19\\
 107 & 20.04$\pm$ 0.01&  56$\pm$ 31&44.17\\
1108 & 20.64$\pm$ 0.04&  54$\pm$ 25&44.16\\
 166 & 21.25$\pm$ 0.01&  55$\pm$ 24&44.15\\
 195 & 21.01$\pm$ 0.05&  52$\pm$ 17&44.14\\
\hline
\multicolumn{4}{c}{Photometric Sample}\\
\hline
 135 & 21.19$\pm$ 0.05& 510$\pm$ 36&45.59\\
1098 & 20.55$\pm$ 0.04& 365$\pm$ 33&45.43\\
 870 & 21.89$\pm$ 0.06& 144$\pm$ 15&45.04\\
1546 & 22.09$\pm$ 0.11&  89$\pm$ 33&44.82\\
1528 & 21.87$\pm$ 0.17&  84$\pm$ 39&44.75\\
1849 & 20.85$\pm$ 0.02& 124$\pm$ 29&44.72\\
1121 & 22.82$\pm$ 0.26&  61$\pm$ 21&44.67\\
 788 & 23.09$\pm$ 0.21&  54$\pm$ 10&44.61\\
 828 & 20.63$\pm$ 0.04& 116$\pm$ 25&44.56\\
 950 & 21.89$\pm$ 0.10&  63$\pm$ 23&44.55\\
 420 & 21.50$\pm$ 0.13&  63$\pm$ 27&44.41\\
 278 & 21.93$\pm$ 0.11&  51$\pm$ 34&44.40\\
1716 & 20.38$\pm$ 0.03&  88$\pm$ 17&44.37\\
1750 & 21.56$\pm$ 0.11&  58$\pm$ 29&44.35\\
1846 & 20.62$\pm$ 0.02&  76$\pm$ 60\tablenotemark{c}&44.30\\
\enddata
\tablenotetext{a}{Galaxy IDs and $K_{s,AB}$ are taken from \citet{Forster06}.}
\tablenotetext{b}{The 24 \micron\ flux errors are estimated within a fixed aperture ($r<5\arcsec$) and are scaled by the ratio of the DAOPHOT PSF fitting flux vs. the fixed aperture flux. The method would overestimate the error for some faint sources.}
\tablenotetext{c}{The source has a low nominal SNR of 1.3 due to the reason mentioned in note b; it has been visually confirmed as a secure detection.} 
\end{deluxetable}